\documentclass[english,12pt]{article}
\usepackage[T1]{fontenc}
\usepackage[latin1]{inputenc}
\usepackage{geometry}
\usepackage{epsfig}
\usepackage{xspace}
\usepackage{cite}

\IfFileExists{url.sty}{\usepackage{url}}
                      {\newcommand{\url}{\texttt}}

\makeatletter

\providecommand{\tabularnewline}{\\}

\newenvironment{lyxlist}[1]
{\begin{list}{}
{\settowidth{\labelwidth}{#1}
 \setlength{\leftmargin}{\labelwidth}
 \addtolength{\leftmargin}{\labelsep}
 }}
{\end{list}}

\newcommand{\psb}{\bar{\psi}}

\newcommand{\eqn}[1]{Eq.(\ref{#1})}
\newcommand{\gev}{\mbox{ GeV}}

\newcommand{\ben}{\begin{enumerate}}
\newcommand{\een}{\end{enumerate}}
\newcommand{\bit}{\begin{itemize}}
\newcommand{\eit}{\end{itemize}}
\newcommand{\bc}{\begin{center}}
\newcommand{\ec}{\end{center}}
\newcommand{\bq}{\begin{equation}}
\newcommand{\eq}{\end{equation}}
\newcommand{\bqa}{\begin{eqnarray}}
\newcommand{\eqa}{\end{eqnarray}}

\newcommand{\nn}{\nonumber}

\newcommand{\plabt}[3]{\raisebox{#3pt}{\epsfig{figure=./#1.eps,width=#2cm}}}
\newcommand{\plaata}[4]{\raisebox{#2pt}{
\epsfig{figure=./#1.eps, width=#3cm,height=#4cm}}}

\def\kperp{\mbox{$k_\perp $\xspace}}

\usepackage{babel}

\makeatother

\date{}

\title{\huge \bf \textsc{Helac-Phegas}: a generator for all parton
level processes
\\[1cm]}

\author{
{\bf Alessandro Cafarella$^{(1)}$\footnote{cafarella@inp.demokritos.gr} ,
Costas G.~Papadopoulos$^{(1)}$\footnote{costas.papadopoulos@cern.ch}} , \\
{\bf Malgorzata Worek$^{(2,3)}$\footnote{worek@particle.uni-karlsruhe.de}}\\
{\normalsize $^{(1)}$Institute of Nuclear Physics, NCSR Demokritos,
15310 Athens, Greece}\\
{\normalsize $^{(2)}$ Institute for Theoretical Physics (ITP),
Karlsruhe University, 76-128 Karlsruhe, Germany}\\
{\normalsize $^{(3)}$ Institute of Physics,
University of Silesia, Uniwersytecka 4, 40-007
Katowice, Poland}
\\[1cm]
\texttt{http://www.cern.ch/helac-phegas}
}

\begin{document}
\maketitle

\vspace{1cm}

\begin{abstract}

The updated version of the \textsc{Helac-Phegas} event generator is presented.
The matrix elements are calculated through Dyson-Schwinger recursive
equations using color connection representation. Phase-space generation
is based on a multichannel approach, including
optimization.   \textsc{Helac-Phegas} generates parton level events with all
necessary information, in the most
recent Les Houches Accord format, for the study of \emph{any
process} within  the Standard Model in hadron and lepton colliders.

\end{abstract}

\newpage
{\bf Program summary}

\vspace{0.5cm}

\begin{small}
\noindent
{\em Program Title:}  \textsc{Helac-Phegas}
                                    \\
{\em Journal Reference:}                                      \\
  %Leave blank, supplied by Elsevier.
{\em Catalogue identifier:}                                   \\
  %Leave blank, supplied by Elsevier.
{\em Licensing provisions:}    none                                \\
  %enter "none" if CPC non-profit use license is sufficient.
{\em Program obtainable from:}
\url{http://helac-phegas.web.cern.ch/helac-phegas/}        \\
{\em Distributed format:}  tar gzip file                                  \\
{\em Programming language:} \textsc{Fortran}                                  \\
{\em Computer:} All                                              \\
  %Computer(s) for which program has been designed.
{\em Operating system:}  Linux                                     \\
  %Operating system(s) for which program has been designed.
{\em Classification:}  11.1, 11.2                                        \\
{\em External routines/libraries:}
Optionally Les Houches Accord (LHA) PDF Interface library. \\
 {\em Nature of problem:}
One of the most striking features of final states in current and future
colliders is the large number of events with several jets.
Being able to predict their features is essential. To achieve this,
the calculations need to describe as accurately as possible the full
matrix elements for the underlying hard processes. Even at leading
order, perturbation theory based on Feynman graphs runs into
computational problems, since the  number of graphs contributing to
the
amplitude grows as $n!$. \\
{\em Solution method:} Recursive algorithms based on Dyson-Schwinger
equations have been developed recently in order to overcome the
computational obstacles.  The calculation of the amplitude, using
Dyson-Schwinger recursive equations,  results in a computational
cost growing asymptotically as $3^n$, where $n$ is the number of
particles involved in the process.  Off-shell subamplitudes are
introduced, for which a recursion relation has been obtained
allowing to express an $n-$particle amplitude in terms of
subamplitudes, with $1-$, $2-$, ... up to $(n-1)$ particles. The
color connection representation is used in order to treat amplitudes
involving colored particles. In the present version
\textsc{Helac-Phegas} can be used to efficiently obtain helicity
amplitudes, total cross sections, parton-level event samples in LHA
format, for arbitrary multiparticle
processes in the Standard Model in leptonic, $p\bar{p}$ and $pp$ collisions.\\
{\em Does the new version supersede the previous version:} yes, partly\\
{\em Reasons for the new version:} substantial improvements, major
functionality upgrade\\
{\em Summary of revisions:} color connection representation, efficient
integration over PDF via the \textsc{Parni} algorithm, interface to LHAPDF,
parton level events generated in the most recent LHA format, $k_\perp$
reweighting for Parton Shower matching,  numerical predictions for amplitudes
for arbitrary processes for phase space points
provided by the user, new user interface and the possibility to run over
computer clusters.\\
{\em Running time:} Depending on the process studied. Usually from seconds
to hours. \\
{\em References:}

[1] A. Kanaki, C. G. Papadopoulos, Comput. Phys. Commun. {\bf 132}
(2000) 306

[2] C. G. Papadopoulos, Comput. Phys. Commun.  {\bf 137} (2001) 247
\end{small}

\newpage
\vspace*{\fill}
\tableofcontents
\vspace*{\fill}
\newpage

\section{Introduction }

Current and forthcoming high energy colliders allow the study of final
states with many hard and well separated jets as well as charged
leptons. These multiparton final states carry the signature of the decay
of massive particles, such as for example W and Z gauge bosons, top quarks,
Higgs bosons, and possibly supersymmetric particles, as well as the effect of
new interactions. Therefore the accurate description of such processes is of
fundamental importance for the study of the properties of the Standard Model
particles as well as for the discovery of new physics.

Multiparton processes can be described within standard perturbation
theory. At the leading order (LO) one needs to calculate  matrix
elements and efficiently integrate them over the available phase
space. Although the procedure is well defined and straightforward in
principle, several challenges are present in practice. First of all,
if the matrix elements are calculated using the standard Feynman
diagrams, the number of which grows factorially with the number of
partons in the process, the set up of an efficient calculational
framework in the multiparton case becomes impossible. The solution
to this problem is by now well known and relies on the use of
recursive equations instead of Feynman graph expansion
\cite{Berends:1987me,Berends:1988yn,Berends:1990ax,Argyres:1992js,
Argyres:1993wz,Caravaglios:1995cd,Draggiotis:1998gr,Caravaglios:1998yr,
Kanaki:2000ey,Kanaki:2000ms,Draggiotis:2002hm,Papadopoulos:2005ky,
Draggiotis:2006er,Duhr:2006iq}.
\textsc{Helac} \cite{Kanaki:2000ey,Kanaki:2000ms} has been the first
implementation of the algorithm based on Dyson-Schwinger equations.
It is able to calculate matrix elements for arbitrary scattering
processes.
A fundamental issue is the
consistent inclusion of all color configurations within QCD for processes with
colored partons. The color connection representation is used in \textsc{Helac},
and the programme is actually able to calculate all possible color connections
for arbitrary processes.
For amplitudes with many colored particles, a Monte-Carlo treatment over
 color should also be used. This issue has also been addressed so far in
references~\cite{Caravaglios:1998yr,Draggiotis:2002hm}. Moreover,
a new version of the
\textsc{Helac} algorithm~\cite{Papadopoulos:2005ky}, based on explicit color
configurations, has been proved very
efficient in  dealing with processes with many colored partons.
%Several techniques have been
%developed so far, and their implementation proved to be quite satisfactory in
%practice \cite{Kleiss:1994qy,Kleiss:1985gy,Ohl:1998jn,
%Draggiotis:2000gm,Hameren:2002tc}.
The next challenge is the phase space integration.
For the phase-space generation, \textsc{Phegas} \cite{Papadopoulos:2000tt}
has been the first implementation of a complete automated algorithm of
multichannel phase space mappings for arbitrary number of external particles.
It uses information generated by \textsc{Helac} and automatically performs
a multichannel phase space generation, utilizing all 'scalarized' Feynman
graphs, for any given process, as phase space mappings.
Finally, since colliding particles like protons are not elementary
one has to take into account a number of elementary parton level
subprocesses.

It should be
mentioned that a number of computer programs, like for example \textsc{Alpgen}
\cite{Mangano:2002ea,Mangano:2001xp}, %\textsc{Ariadne}
\cite{Lonnblad:1992tz}, \textsc{MadEvent}
\cite{Stelzer:1994ta,Maltoni:2002qb,Alwall:2007st} \textsc{Sherpa}
\cite{Krauss:2001iv,Gleisberg:2003xi}, \textsc{Whizard} \cite{Kilian:2007gr}
are available, which implement some of these techniques and can be used to
obtain leading order predictions for processes with a rather high number of
final state particles.

In the present paper we will summarize the new developments in the
\textsc{Helac-Phegas} which fall into two categories:
\begin{enumerate}
\item {\bf Functionality:}
\begin{enumerate}
\item Full implementation of QCD in the color connection representation
\item Automatic summation over subprocesses
for $pp$ and $p\bar{p}$ colliders and $e^+e^-$ colliders
\item Built-in CTEQ6l1 PDF support \cite{Pumplin:2002vw,Stump:2003yu}
with alternative interface to the LHAPDF library \cite{Whalley:2005nh}
\item $k_T$ reweighting for matching algorithm to parton shower
programs, e.g. \textsc{Pythia} \cite{Sjostrand:2006za,Sjostrand:2007gs},
 \textsc{Herwig}\cite{Corcella:2000bw} or \textsc{Herwig++}\cite{Bahr:2008pv}
\item Parton-level event files in the LHA format \cite{Boos:2001cv,Alwall:2006yp}
\item Extended set of standard cuts
\end{enumerate}
\item {\bf User interface:}
\begin{enumerate}
\item Program control using a single user input file managed by scripts
\item Easy access to numerical amplitude values for a user given phase space
point
\item Set up for parallel runs on computer clusters
\end{enumerate}
\end{enumerate}

The present paper is organized as follows. In the next section, after a brief
summary of the \textsc{Phegas} phase space generator and the \textsc{Helac}
 algorithm, we describe the treatment of the color degrees of freedom, based
 on the color connection representation
and the matching to the parton shower algorithms. The rest of the
publication describes in detail the usage of the program. We
summarize with a short outlook.

\section{The \textsc{Helac-Phegas} Algorithm}

The core of \textsc{Helac-Phegas} is made of a phase space generator and a
matrix element evaluator. The algorithms of both have already been described
previously in \cite{Papadopoulos:2000tt} and \cite{Kanaki:2000ey}
respectively. Therefore,
we will only give here a very short overview of the main ideas.

%%%%%%%%%%%%%%%%%%%%%%%%%%%%%%%%%%%%%%%%%%%%%%%%%%%%%%%%%%%%%%%%%%%%%%%%%%%%%%%%
\subsection{\textsc{Phegas} - Phase Space Generation}
%%%%%%%%%%%%%%%%%%%%%%%%%%%%%%%%%%%%%%%%%%%%%%%%%%%%%%%%%%%%%%%%%%%%%%%%%%%%%%%%
\label{phegas} \textsc{Phegas} \cite{Papadopoulos:2000tt}, is a
multichannel self optimizing phase space generator. It automatically
constructs all possible phase-space mappings in order to optimally
describe the peaking structures of a given scattering process. These
phase-space mappings, called also \emph{channels}, have a one-to-one
correspondence to the Feynman diagrams contributing to the given
process. The information concerning the pole structure of all
Feynman diagrams is produced by \textsc{Helac} by
using the skeleton construction of the solution
of the Dyson-Schwinger recursive equations. The self-optimization is performed
according to the ideas presented in~\cite{Kleiss:1994qy}. Moreover
since in several practical applications the number of phase-space
channels may become quite large (of the order of many thousands),
optimization may be used to significantly reduce this number, based
on the contribution of the corresponding channels
to the total variance. At the end a few tens of channels are kept only,
which is usually sufficient to give an
adequate phase space efficiency. Let us stress that the selection of
these channels is done automatically by the optimization procedure.

In the case of $pp$ and $p\bar{p}$ collisions, besides phase space generation
for a given center-of-mass energy, an integration over the fractions $x_1$ and
$x_2$ of the momenta of the initial partons, weighted by the  parton
distribution functions, is necessary. This integration is optimized via
the \textsc{Parni} algorithm \cite{vanHameren:2007pt}. Its efficiency has been
 proven to be very good.

%%%%%%%%%%%%%%%%%%%%%%%%%%%%%%%%%%%%%%%%%%%%%%%%%%%%%%%%%%%%%%%%%%%%%%%%%%%%%%%%
\subsection{\textsc{Helac} - Amplitude Computation}
%%%%%%%%%%%%%%%%%%%%%%%%%%%%%%%%%%%%%%%%%%%%%%%%%%%%%%%%%%%%%%%%%%%%%%%%%%%%%%%%

\textsc{Helac} uses an
alternative~\cite{Caravaglios:1995cd,Draggiotis:1998gr,Caravaglios:1998yr,
Draggiotis:2002hm} to the Feynman graph representation of
the scattering amplitude which is provided by the
Dyson-Schwinger approach \cite{Kanaki:2000ey}.

Dyson-Schwinger equations express recursively the
$n$-point Green's functions in terms of the $1-$, $2-$, $\ldots$, $(n-1)$-point
functions. For instance in QED these equations can be written as follows:

\[
\plaata{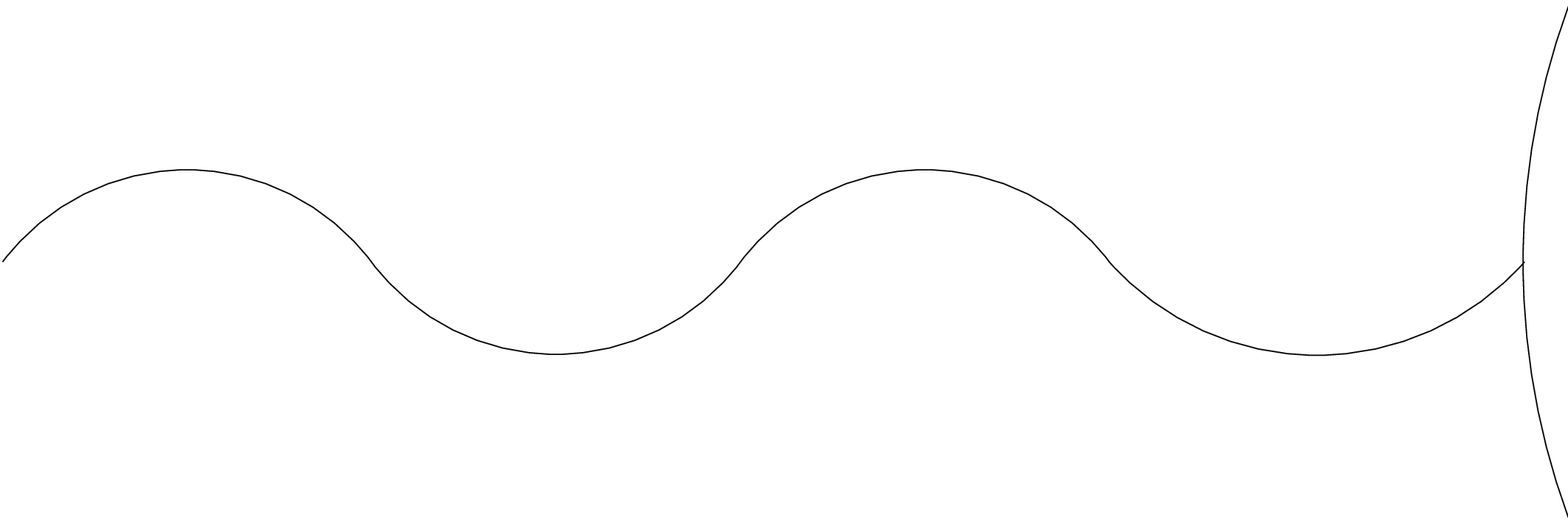}{-15}{1.8}{1}\;\;=\;\plaata{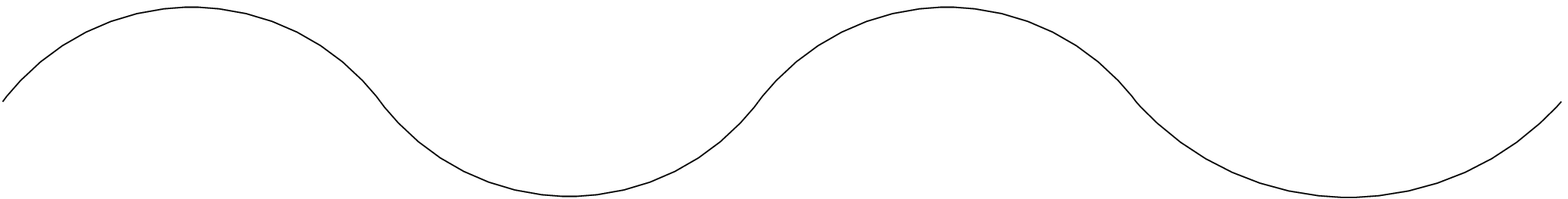}{0}{1.8}{0.2}\;
+\;\plaata{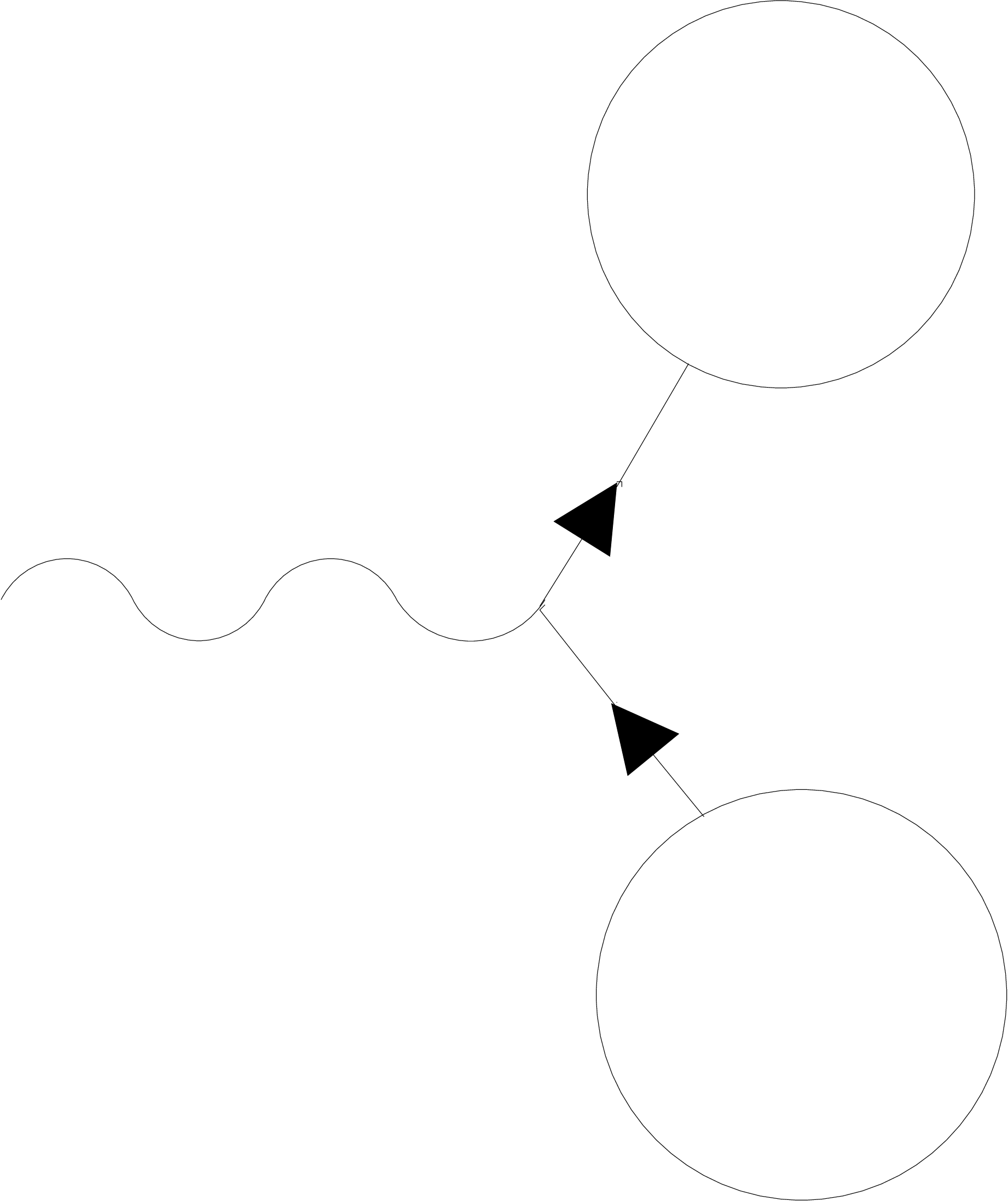}{-25}{1.8}{2}
\]
%\\[6pt]
\bqa && b^\mu(P)=\sum_{i=1}^n \delta_{P=p_i} b^\mu(p_i) \nn
\\
&&+ \sum_{P=P_1+P_2} (ig)\Pi^\mu_\nu\; \psb(P_2)\gamma^\nu\psi(P_1)
\epsilon(P_1,P_2) \nn \eqa where
\[
 b_\mu(P) \;=\; \plaata{bos1}{-5}{0.8}{0.5} \;\;\;\;
   \psi(P)  \;=\; \plaata{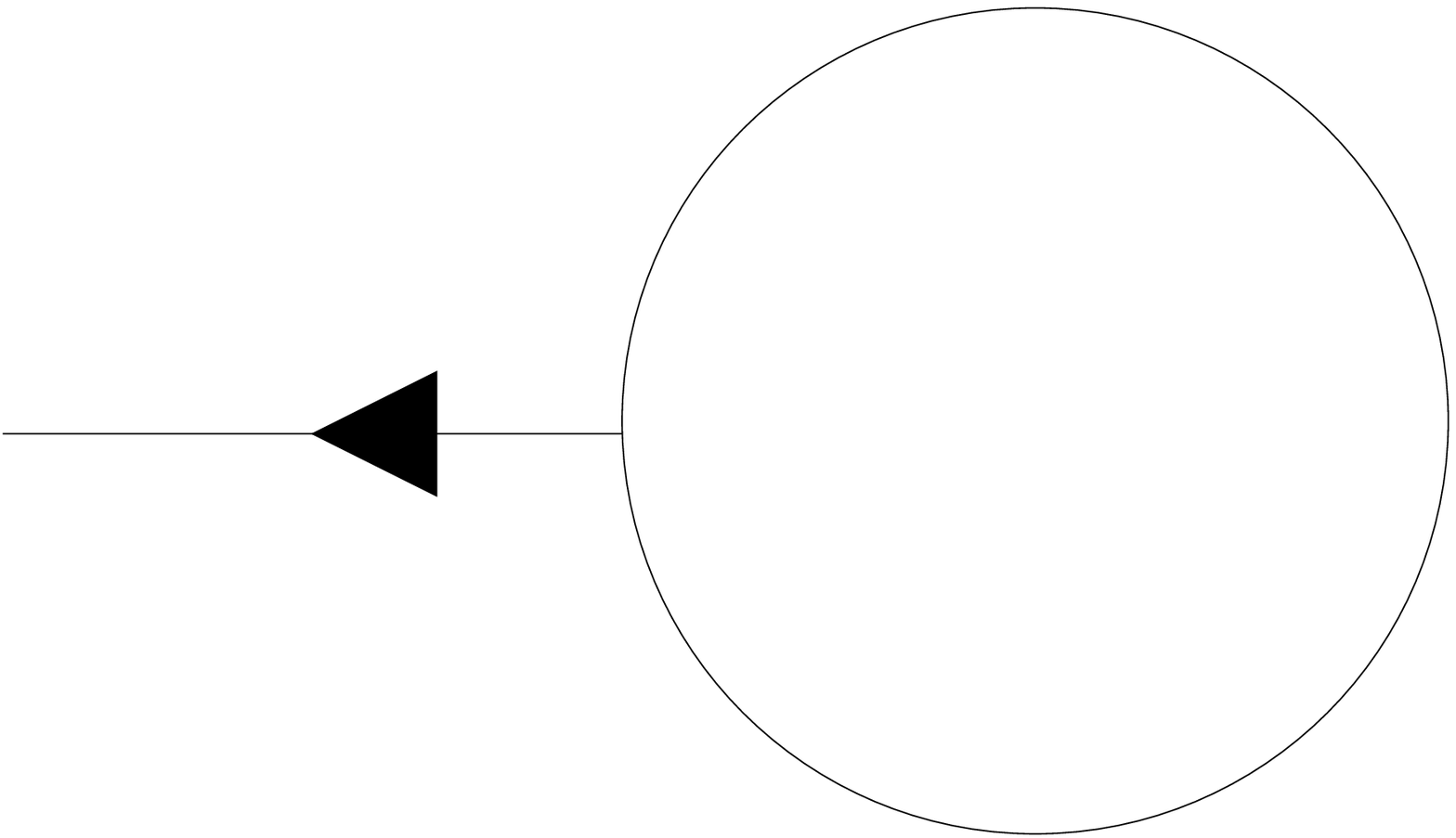}{-6}{0.8}{0.6}\; \;\;\;
   \psb(P)  \;=\; \plaata{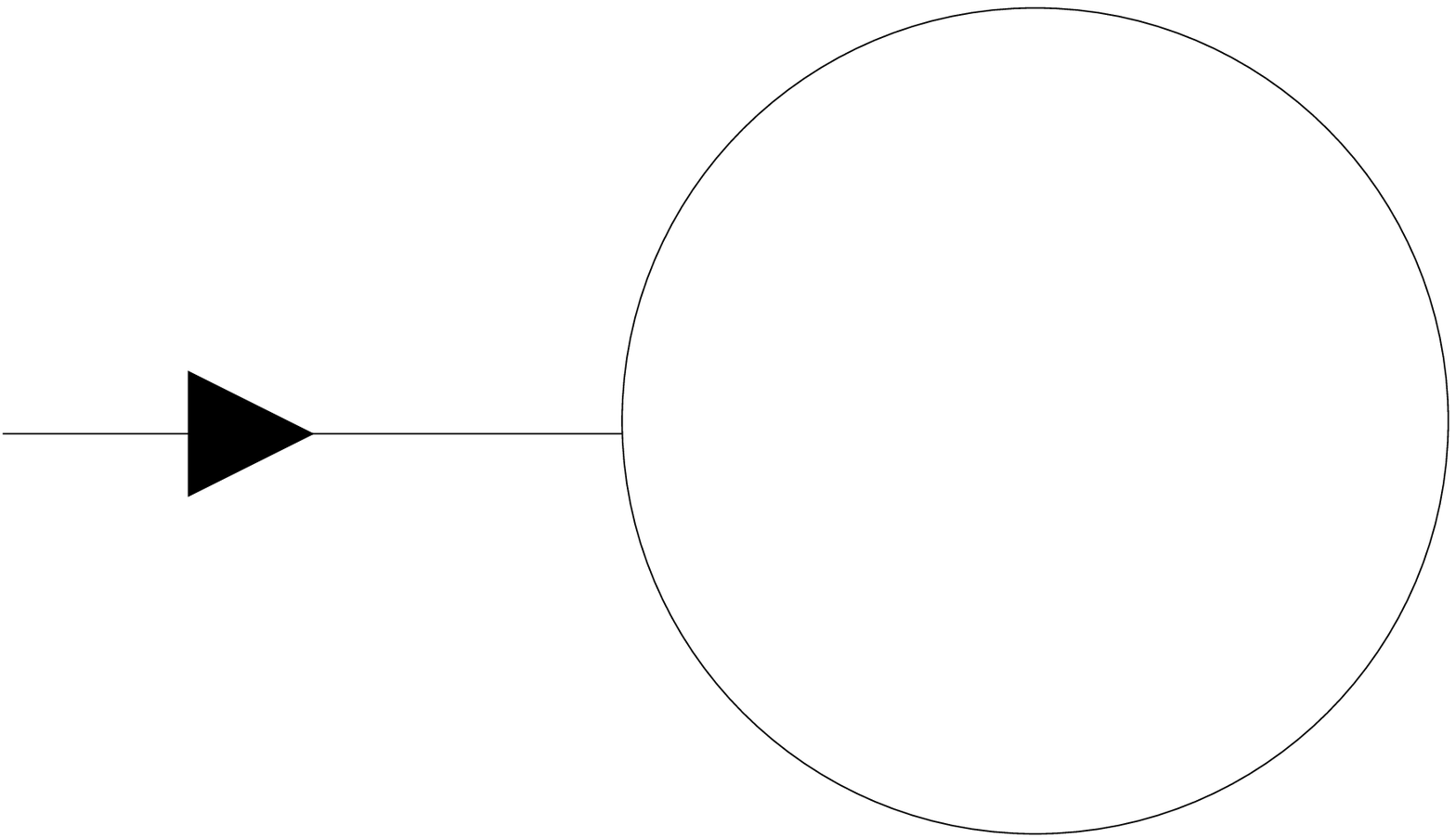}{-6}{0.8}{0.6}
\] describes a generic $n$-point Green's function with respectively one
outgoing photon, fermion or anti\-fermion leg carrying momentum $P$.
$\Pi_{\mu\nu}$ stands for the boson propagator and $\epsilon$ takes
into account the sign due to fermion antisymmetrization.

The computational cost of \textsc{Helac-Phegas}'s
algorithm grows as $\sim 3^n$, which
essentially counts the steps used to solve the recursive equations.  Obviously
for large $n$ there is a tremendous saving of computational time, compared to
the $n!$ growth of the Feynman graph approach.

\textsc{Helac-Phegas} is using a \emph{top-down approach} in solving the
recursive equations.  In the first phase, a skeleton is constructed, composed
by a series of vertices that are participating in the given amplitude. All
this information is encoded using integer arithmetic. Then this information is
used in order to actually calculate the amplitude. In this second phase,
momenta from \textsc{Phegas} are used to compute wave functions.
Based on  the
skeleton information the subamplitudes are computed using floating
arithmetic. The result is the value of the matrix element for each given
helicity and color connection configuration.

%%%%%%%%%%%%%%%%%%%%%%%%%%%%%%%%%%%%%%%%%%%%%%%%%%%%%%%%%%%%%%%%%%%%%%%%%%%%%%%%
\subsection{Color representation}
%%%%%%%%%%%%%%%%%%%%%%%%%%%%%%%%%%%%%%%%%%%%%%%%%%%%%%%%%%%%%%%%%%%%%%%%%%%%%%%
%
The treatment of the color degrees of freedom is a very important
issue, especially for amplitudes involving many colored particles.
In the most commonly used approach we can express a $n-$gluon
amplitude in terms of the so-called color-ordered amplitudes, using
the following equation,
\begin{equation}\label{sun}
    {\cal M}^{a_1 a_2 \ldots a_n}=\sum_{\sigma} Tr(T^{a_{\sigma_1}}
\ldots T^{a_{\sigma_n}})
    {\cal A}_\sigma
\end{equation}
$\sigma=(\sigma_1,\sigma_2,\ldots,\sigma_n)$ is permutation of
$(1,2,\ldots,n)$. The sum is extended over all $(n-1)!$
permutations, where ${\cal A_\sigma}$ are functions of momenta and
helicities of the external particles. The important thing is that
these ${\cal A_\sigma}$ color-ordered amplitudes can be calculated
using the so-called color-ordered Feynman rules. In the framework of
the Berends-Giele recursive equations, which is the Dyson-Schwinger
equations subject to the restriction of ordering of the external
particles (according to the permutation $\sigma$), the computational
cost grows like $n^4$, exhibiting a polynomial growth.
When calculating the full color-summed matrix element
squared, all $(n-1)!$ terms have to be computed.
This approach is some times referred as the \emph{$SU(N_c)$ approach}.
The inclusion of quark-antiquark pairs is possible, but the analogue
of \eqn{sun} becomes more complicated.

The \emph{color connection representation} is based on a \emph{$U(N_c)$
representation} of the fields. In that case amplitudes involving quarks
and gluons can  by described in a unified way. To this end all indices of
the adjoint representation are  transformed to indices in the
fundamental representation by multiplying with $t^a_{i,j}$,
\begin{equation}
    {\cal M}^{i_1 i_2 \ldots i_n}_{j_1 j_2 \ldots j_n}=
    \sum_{a}{\cal M}^{a_1 a_2 \ldots a_n}t^{a_1}_{i_1 j_1} \ldots t^{a_n}_{i_n j_n}.
\end{equation}
When quarks are only involved no action is required. The $U(N_c)$
representation has been introduced in Ref. \cite{'tHooft:1973jz}. It has been
used in connection with matrix element evaluation in
\textsc{Helac-Phegas} \cite{Kanaki:2000ey,Kanaki:2000ms}.
A Lagrangian based approach has also been used in \cite{Maltoni:2002mq}.

Now the amplitudes can be written as follows:
\begin{equation}\label{un}
    {\cal M}^{i_1 i_2 \ldots i_n}_{j_1 j_2 \ldots j_n}=
    \sum_{\sigma} \delta_{{i_{\sigma_1}} j_1} \delta_{{i_{\sigma_2}} j_2}
\ldots \delta_{{i_{\sigma_n}} j_n} {\cal A_\sigma}.
\end{equation}
In the above expression we have chosen to fix the order of  anti-color
indices.  Without altering the result, the same can be done with the color
indices. We have kept the same
notation for the functions ${\cal A_\sigma}$ in \eqn{sun} and
\eqn{un}. This is indeed the case for the $n-$gluon amplitude but
not when quarks are involved. In this representation there is no
distinction anymore between gluons and quarks. Gluons are
represented by a pair of color-anti-color lines, where quarks
(anti-quarks) are represented by color (anti-color) lines. The number
of permutations is in principle equal to $n_l!$, where $n_l$ is the
number of color (anti-color) lines. In a process with $n_g$ gluons
and $n_q$ quarks (and $n_q$ anti-quarks), $n_l=n_g+n_q$.

The important thing is again that the ${\cal A_\sigma}$ function can be
calculated using a given set of Feynman rules.
For instance, the three-gluon vertex can be rewritten as
\begin{equation}
\sum f^{abc} t^a_{AB} t^b_{CD} t^c_{EF}=-\frac{i}{4} (
\delta_{AD}\delta_{CF}\delta_{EB}-
\delta_{AF}\delta_{CB}\delta_{ED})
\end{equation}
where on the right hand side only products of $\delta$'s appear. Let
us introduce a more compact notation and associate to each gluon a
label $(i,\sigma_i)$, which refers to the corresponding color index
of the previous equation, namely $1\to A$, $\sigma_1\to B$ and so
on. The use of $\sigma_i$ labels will become clear in a moment. With
this notation, the first term of the above equation is proportional
to
\begin{equation}
\delta_{1\sigma_2}\delta_{2\sigma_3}\delta_{3\sigma_1}
\end{equation}
and its diagrammatic representation is given by
\[
\plabt{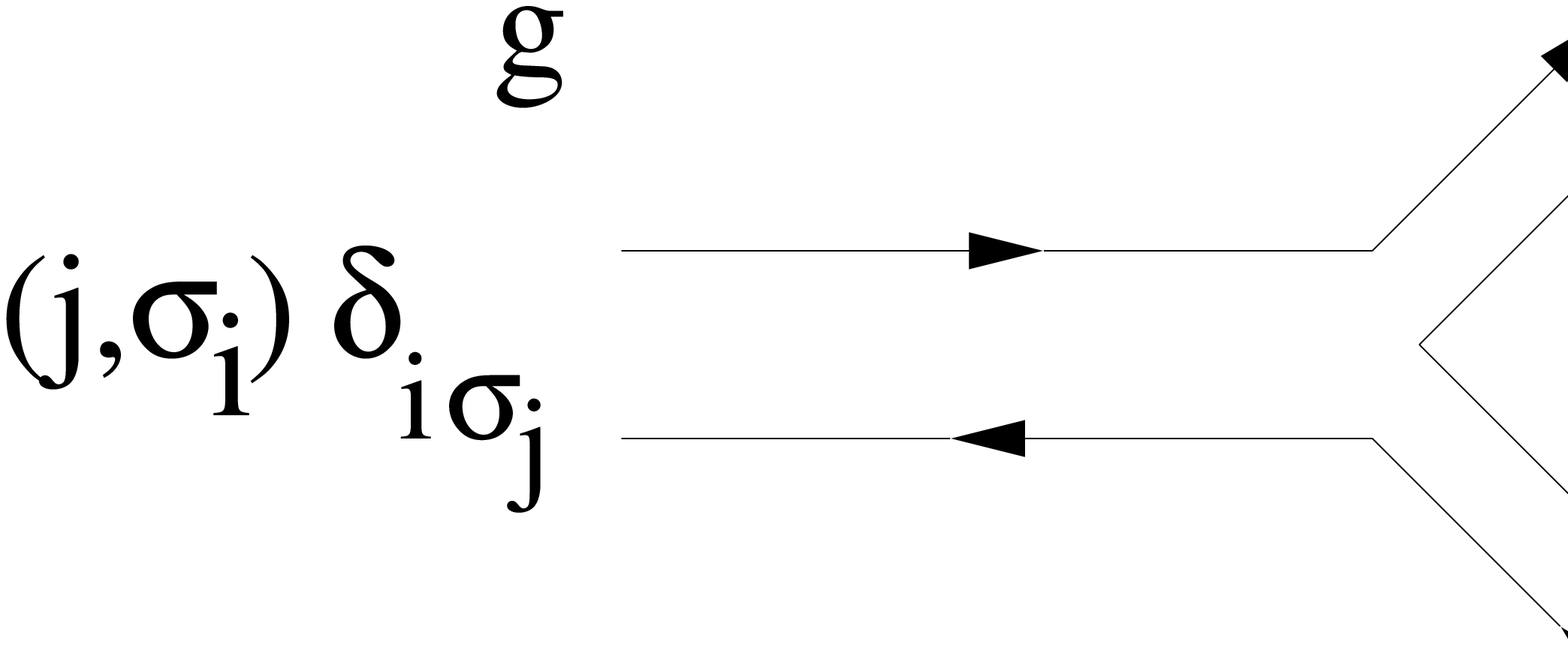}{6}{0}
\]
%For the same graph with inverted arrows, a minus sign and
%interchanged $i\leftrightarrow j$ have to be included as well.
The
momentum part of the vertex is still the usual one,
\begin{equation}
g_{12} \left(p_1-p_2\right)^3+g_{23}   \left(p_2-p_3\right)^1+g_{31}
\left(p_3-p_1\right)^2
\end{equation}
where $g_{12}$ stands for $g_{\mu_1 \mu_2}$, etc.
For the $q\bar{q}g$ vertex we have
\begin{equation}
\sum t^a_{AB} t^a_{CD} =\frac{1}{2} ( \delta_{AD}\delta_{CB}-
\frac{1}{N_c}\delta_{AB}\delta_{CD})
\end{equation}
and if we associate a label $(i,0)$($(0,\sigma_i)$) to a
quark(antiquark) we have two possible vertices
\[
\plabt{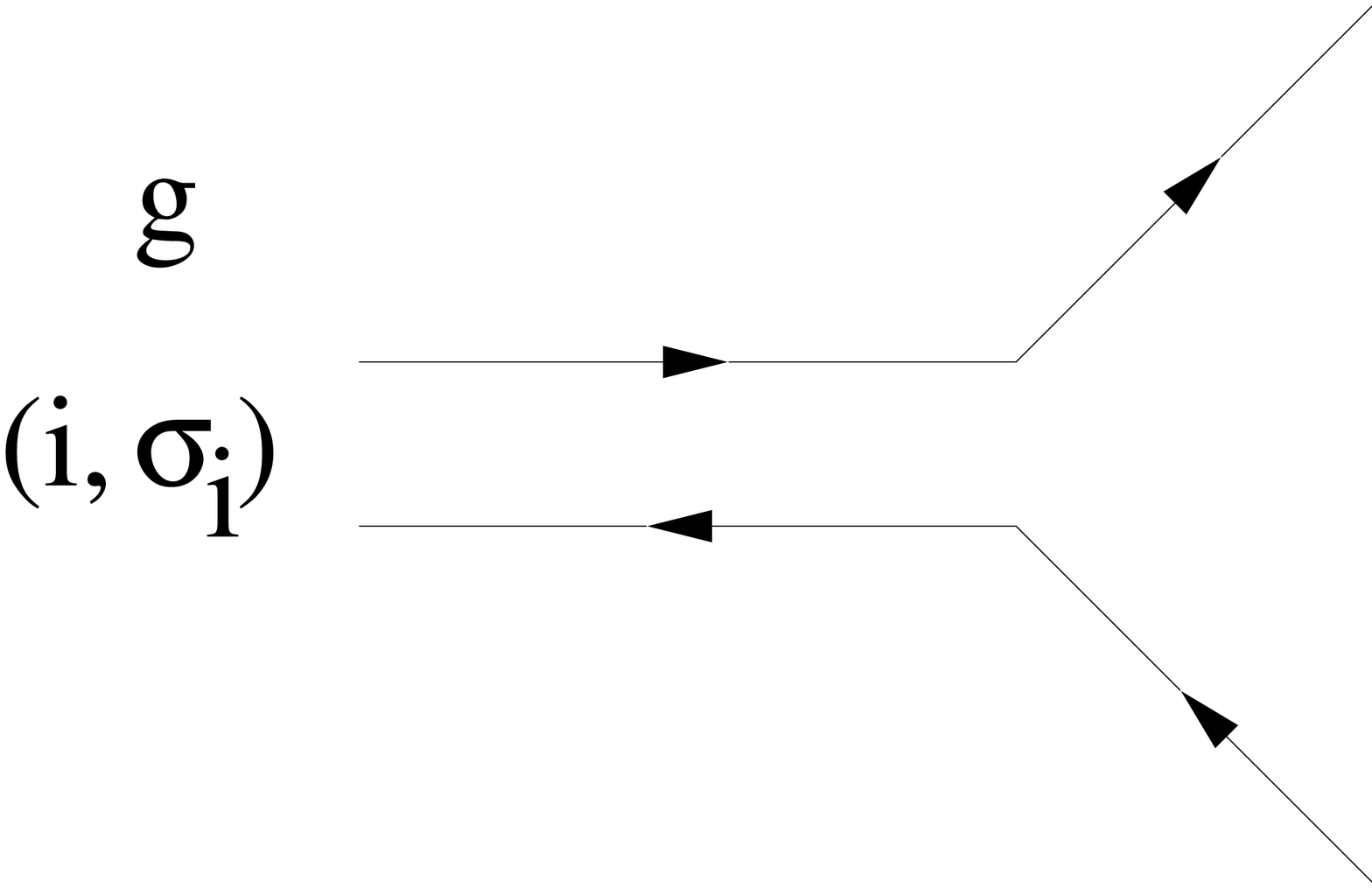}{6}{0}
\]
as well as the one with the 'neutral' gluon, namely,
\[
\plabt{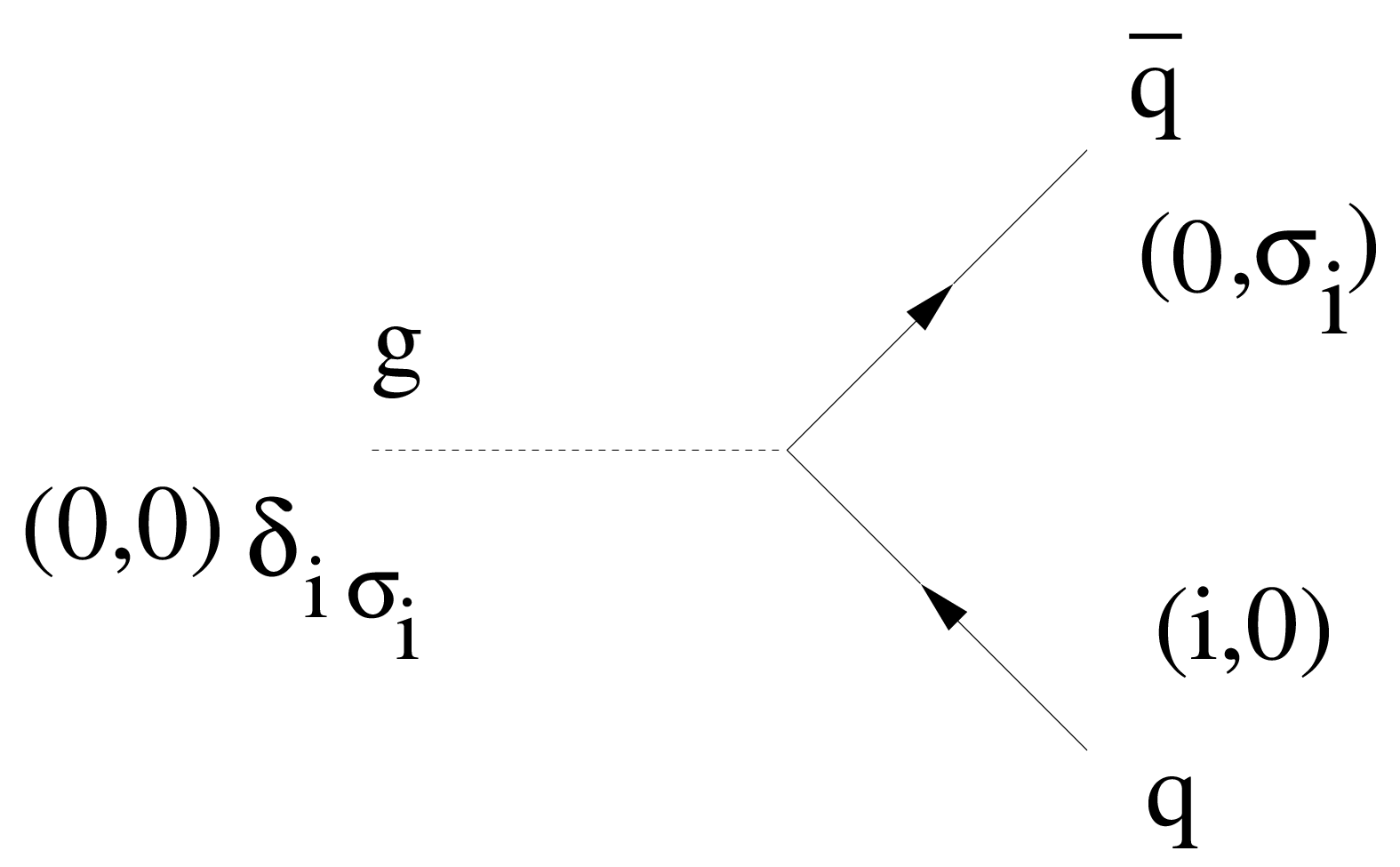}{6}{0}
\]
the last one with an extra factor proportional to $i/\sqrt{N_c}$.
Finally the four-gluon vertex is given by a color factor
proportional to
\begin{equation}
\delta_{1\sigma_3} \delta_{3\sigma_2} \delta_{2\sigma_4} \delta_{4\sigma_1}
\end{equation}
and a Lorentz part
\begin{equation}
2 g_{12} g_{34} - g_{13} g_{24} - g_{14} g_{23}
\end{equation}
\[
\plabt{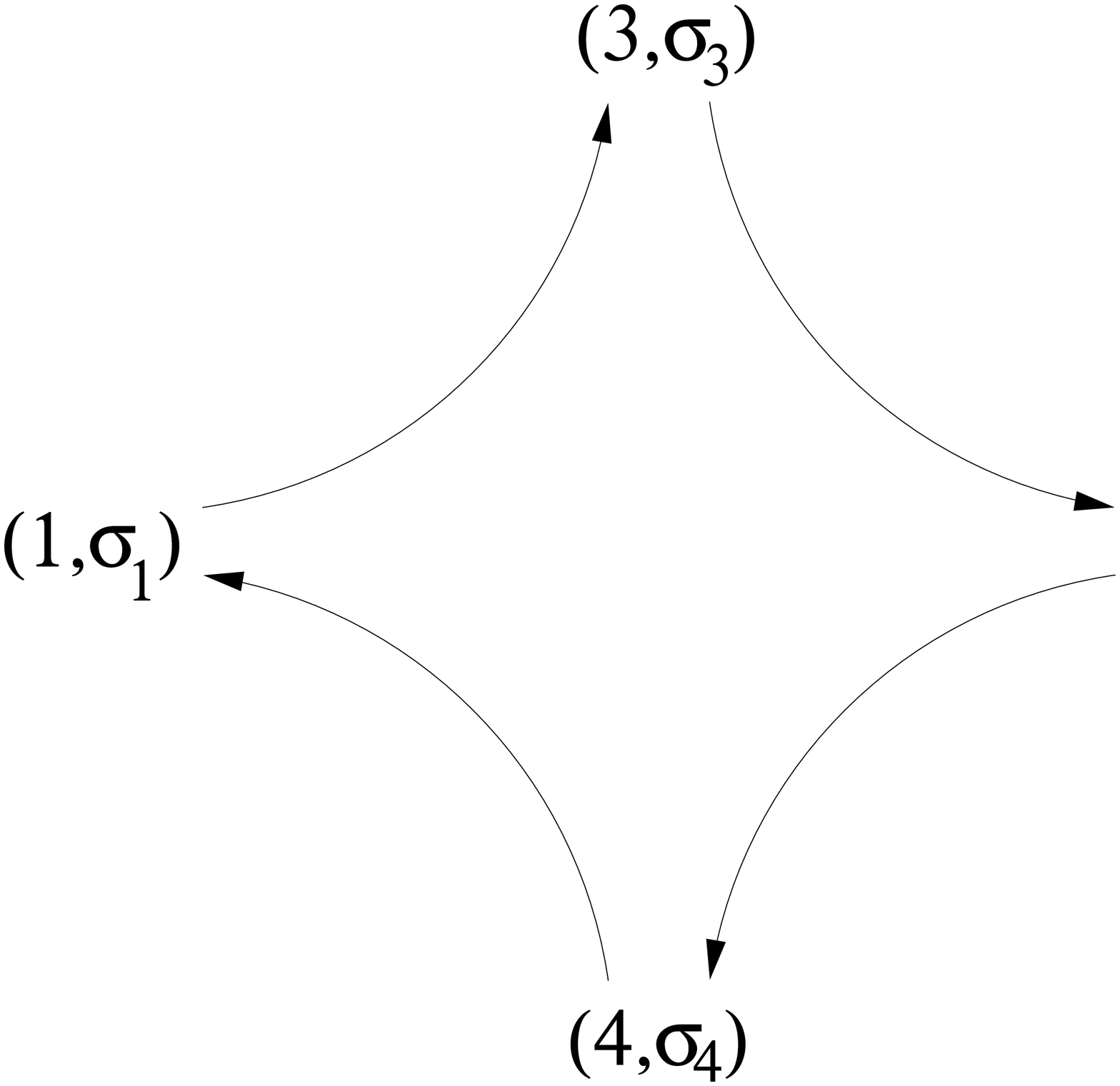}{6}{0}
\]
%Of course by keeping the $(1,\sigma_1)$ fixed in the above figure,
%there are six permutations among $(2,\sigma_2), (3,\sigma_3), (4,\sigma_4)$
%that are possible (topologically inequivalent) in general;
%a given assignment of $\sigma_2,\sigma_3,\sigma_4$ will peak one
%of them with the corresponding lorentz structure.

To make use of the color representation described so far, let us
assign  a label $(i,\sigma_I(i))$ to each external gluon, $(i,0)$ to
a (incoming) quark and $(0,\sigma_I(i))$ to an (incoming) antiquark,
where $i=1\ldots n_l$ and $\sigma_I, I=1\ldots n_l!$ being a
permutation of $\{1\ldots n_l\}$. It is clear that since all
elementary color factors appearing in the color decomposition of the
vertices are proportional to $\delta$-symbols, the total color
factor can only be given by
\begin{equation}
{\cal F}_I=\delta_{1\sigma_I(1)}\delta_{2\sigma_I(2)}
\ldots\delta_{n\sigma_I(n)}
\end{equation}
Moreover the color matrix, defined as
\begin{equation}
\label{colormatrix}
{\cal C}_{IJ} = \sum_{\mbox{\scriptsize  colours}} {\cal F}_I {\cal
F}_J^\dagger
\end{equation}
with the summation running over all colors, $1\ldots N_c$, has a
very simple representation
\begin{equation}
{\cal C}_{IJ} = \left( N_c \right)^{m(\sigma_I,\sigma_J)}
\end{equation}
where $1\le m(\sigma_I,\sigma_J)\le n$ counts the number of cycles
made by the elements of the permutations $\sigma_I$ and $\sigma_J$.

The practical implementation of these ideas is rather straightforward. Given
the information on the external particles contributing to the process, we
associate color labels of the form $(i,\sigma_i)$, depending on their
flavor. The Feynman rules to build up the higher level
subamplitudes, are the ones described above. The result, for each
permutation, is a new colored ordered ${\cal A}$-function that corresponds to
the given color factor ${\cal F}_I$. The computational cost is again given,
as in the case of Berends-Giele recursive equations, by $n^4$.
The total color-summed squared amplitude is obtained after summing over
all $n_l!$ color connection configurations using the color matrix ${\cal C}$.

In the generation of unweighted events a color connection
information is needed for a proper matching with parton shower
algorithms. For the selection of the color connection structure to
be assigned to a given event we consider only those ${\cal
A}$-functions that correspond to color connections without a
'neutral' gluon involved. The choice is based then on a probability
distribution defined by
\[
P_I=\frac{|{\cal A}_I|^2}{\sum_J |{\cal A}_J|^2}, \,\,\,\,\, P_I>0,
\,\,\,\,\, \sum_I P_I=1
\]
where $I,J$ run over all color connections considered.

%%%%%%%%%%%%%%%%%%%%%%%%%%%%%%%%%%%%%%%%%%%%%%%%%%%%%%%%%%%%%%%%%%%%%%%%%%%%%%%%
\subsection{Interfacing and matching to parton shower programs}
\label{matching}
%%%%%%%%%%%%%%%%%%%%%%%%%%%%%%%%%%%%%%%%%%%%%%%%%%%%%%%%%%%%%%%%%%%%%%%%%%%%%%%%

\textsc{Helac-Phegas} generates a Les Houches Accord (LHA) file
\cite{Boos:2001cv,Alwall:2006yp} in a completely automatic way, with
all the necessary information needed to interface to
 \textsc{Pythia} and/or \textsc{Herwig} parton shower
and hadronization programs.

Possible problems of  double counting
of jets may arise when interfacing fixed order tree level matrix
elements to parton shower codes. The reason is
that a jet can appear both from relatively hard emission during shower
evolution or from inclusion of higher order matrix elements.  In
order to deal with this problem a matching or merging algorithm
has to be applied. It removes double counting of jet configurations
and provides a smooth transition between the part of the phase space
covered by parton showers and the one described by matrix elements.
We refer to \cite{Catani:2001cc,Krauss:2002up,
Schalicke:2005nv,Lavesson:2005xu,Mangano:2006rw,Alwall:2007fs} for a
detailed account of these algorithms.
Here we will briefly present what has been incorporated in
\textsc{Helac-Phegas},  which is the so-called $k_{\perp}$ reweighting
algorithm based on the $k_{\perp}$ measure \cite{Catani:1992zp,Catani:1993hr}.
Moreover, we will also give a brief description of the whole framework,
within the so-called MLM matching approach~\cite{Mangano:2006rw}, which goes
beyond \textsc{Helac-Phegas} capabilities and relies on the interface with
a parton shower algorithm.

The matching algorithm can be described as follows:
\def \et {\ensuremath{E_T}}
\def    \dRjj             {\mbox{$\Delta R_{\sss{jj}}$}}

\def    \be             {\begin{equation}}
\def    \ee             {\end{equation}}
\def \sss {\scriptscriptstyle}
\def    \ptpart             {\mbox{$p^{\sss{part}}_{\sss T}$}}
\def    \etmin        {\mbox{$E_T^{\sss{min}}$}}
\def    \ptmin             {\mbox{$p_{\sss T}^{\sss{min}}$}}
\def    \etapart             {\mbox{$\eta_{\sss{part}}$}}
\def    \etamax         {\mbox{$\eta_{\sss{max}}$}}
\def    \as             {\ifmmode \alpha_s \else $\alpha_s$ \fi}
\def    \qzero            {{\mbox{$Q_0$}}}
\def    \qzerosq            {{\mbox{$Q^2_0$}}}
\def \oatwo {\mbox{$ {\cal O} (\alpha_s^2)$}}
\def    \etaclmax       {\mbox{$\eta^{\mathrm clus}_{\mathrm max}$}}
\def    \rmin           {\mbox{$R_{\sss{min}}$}}
\def    \rclus          {\mbox{$R_{\mathrm{clus}}$}}
\def    \drpp           {\mbox{$\Delta R_{\mathrm{pp}}$}}
\def    \ptsq           {\mbox{$p_{\mathrm T}^2$}}
\def    \pt           {\mbox{$p_{\mathrm T}$}}
\def    \etaclmax       {\mbox{$\eta^{\mathrm clus}_{\mathrm max}$}}
\def    \etclus         {\mbox{$E^{\mathrm{clus}}_{\mathrm T}$}}
\def    \gev            {\mbox{$\mathrm{GeV}$}}

\begin{enumerate}
\item

At the beginning one generates all parton level configurations for
all final state parton multiplicities $n$ up to a given $N$.
Due to the presence of collinear and soft singularities
one has to use a set of \emph{parton level} kinematical cuts:

  \begin{equation} \label{eq:cuts}
    \ptpart>\ptmin\; , \quad \vert \etapart \vert < \etamax \; , \quad
    \dRjj>\rmin \; ,
  \end{equation}

  where \ptpart\ and \etapart\ are the transverse momentum and
  pseudorapidity of the final state partons, and \dRjj\ is their minimal
  separation in the $(\eta,\phi)$ plane.  The parameters $\ptmin$,
  $\etamax$ and $\rmin$ are called generation parameters, and are the
  same for all $n=1,\,\dots,\,N$.

\item

During the generation of  events, the renormalization scale (running
$\alpha_s$) is set according to the CKKW
prescription\cite{Catani:2001cc,Krauss:2002up}. To this end a tree
branching structure is defined for each event, allowing, however,
only for those branchings which are consistent with the color
structure of the event. In \textsc{Helac-Phegas} this is  inherent
in the matrix element calculation.  More precisely for a pair of
final-state partons $i$\/ and $j$, we use the \kperp-measure defined
by

  \[ k_{\perp,ij}=\Delta R_{ij}\, {\mathrm{min}}(\pt_i,\pt_j) \; , \nonumber
  \]

  where $\Delta R_{ij} =\sqrt{\Delta\eta_{ij}^2+\Delta\phi_{ij}^2}$,
while for a pair of
initial/final state partons we have $k_\perp^2=\ptsq$, i.e.
the $\pt$ of the final state parton.

Then if $k^2_{\perp,ij}$ is minimal the 'partons' are clustered, and
the resulting 'parton' is again classified as a final state parton
with $p=p_i+p_j$ and adjusted flavor and color flow. In the case
when incoming and outgoing partons are clustered, the new 'parton'
is considered as incoming and its momentum is given by $p=p_j-p_i$.
As a result we obtain a chain of \kperp-measures. For every node, a
factor of $\alpha_s(k^2_{\perp,\rm node})/\alpha_s(Q^2_{\rm hard})$
is multiplied into the weight of the event. For the un-clustered
vertices as well as for the scale used in the parton density
functions, the hard scale of the process $Q_{\rm hard}^2$ is used.
No Sudakov reweighting is applied.

\item
At this level, an LHA file has been created, which can be processed
independently for showering and hadronization. In that sense
\textsc{Helac-Phegas} involvement ends here. In order to have a
self-contained presentation for the reader, we will now give a brief
description of a possible matching algorithm related to the
parton-showered events we have already used in
Ref.\cite{Alwall:2007fs}. For all partons resulting from the shower
evolution a jet cone algorithm is applied. The resulting jets are
defined by $E_{T \rm min}^{\rm clus}$, $\eta_{\rm max}^{\rm clus}$
and by a jet cone size $R_{\rm clus}$. The parton from the
parton-level event is then associated to one of the constructed
jets. Starting from the parton with the highest $p_T$ we select the
closest jet ($1.5\,\times\,R_{\rm clus}$) in the
pseudo-rapidity/azimuthal-angle space. All subsequent partons are
matched iteratively to jets. If this is impossible, the event is
rejected. Additionally, for $n<N$, matched events with the number of
jets greater than $n$ are rejected, whereas for $n=N$, i.e.
the highest multiplicity, events with extra jets are kept, only if
they are softer than the $N$  matched jets. This procedure provides
the complete inclusive sample. The harder the threshold for the
transverse energy of the jets used in the matching, \etclus, the
fewer the events rejected by the extra-jet veto (i.e. smaller
Sudakov suppression), which means that the role given to the shower
approximation in the production of jets is more dominant. On the
other hand using a lower threshold the role of the matrix-element
calculation is enhanced, which can be seen as equivalent to a more
substantial Sudakov suppression, reducing thus the role of the
parton shower algorithm in the production of the inclusive sample.

\end{enumerate}

In the current version, \textsc{Helac-Phegas} 1.2.0, the
$\kperp$-reweight algorithm described so far is included and it can
be applied (optionally) to all processes. Moreover, the hard scale
of the process can be defined by the user in the
\texttt{getqcdscale.h} file as described in the next section.

\section{Running the code}

\subsection{Description of the code}
\label{run}

The basic structure of the program is already described in
\cite{Kanaki:2000ey}. Just to summarize,  the program reads several input
variables and in the so called \emph{first phase}\footnote{By default both
first and second phases are executed in one run.} the solution to the
Dyson-Schwinger equations is constructed. In contrast to other programs
%like \textsc{MadGraph} or \textsc{Alpgen}
all necessary information is kept in
memory, no extra subdirectories are generated or needed\footnote{There is,
however, the option to keep this information in a file and use it later on as
well.}. This is the reason why \textsc{Helac-Phegas} has a
 really compact and small
source directory.  All arithmetics in the \emph{first phase} is completely
integer and results in the construction of what we call a skeleton, a solution
of the recursive equations.
In the \emph{second phase} based on the information produced so far,
the program generates \emph{completely automatically} phase space points
by  \textsc{Phegas}, and then calculates the
corresponding matrix elements, for each color connection configuration
separately.
In this way not only the total weight can be computed but also the
weight of each color connection configuration is available for free. Color and
phase space un-weighting is therefore straightforward.

\textsc{Helac-Phegas} incorporates
all Standard Model particles and couplings in both
unitary and Feynman gauges. Unstable particles are treated in a fully
consistent way, either by including a fixed width or by using the complex mass
scheme~\cite{Argyres:1995ym,Beenakker:1996kn,Denner:2006ic}.
Moreover, when
final states are considered, all correlations (spin, color) are automatically
taken into account with no approximation involved.

\subsection{Preliminary operations}

Unpacking the code with the command\texttt{}~\\
\texttt{tar zxvf helac\_1.2.0.tar.gz}~\\
will create the directory \texttt{helac\_1.2.0}.

The program makes use of the Les Houches Accord PDF Interface library
(LHAPDF). If this library is not installed
on your machine, download it from
 \url{http://projects.hepforge.org/lhapdf/}
and follow the installation instructions\footnote{Note that if you do not
have root privileges, you can install the library in a local directory.}.

After that you have to edit once for all the file \texttt{myenv}.
Here you have to set up the following environmental variables:

\begin{lyxlist}{00.00.0000}
\item [{\texttt{FC}}] is your \textsc{Fortran} compiler;
nowadays \texttt{gfortran} is
freely available and can be downloaded from
\texttt{http://gcc.gnu.org/wiki/GFortran}. The code has been also tested with
other compilers, like \texttt{Lahey Fujitsu} \texttt{lf95},
\texttt{Intel Fortran} \texttt{ifort} and \texttt{GNU}
\texttt{g95}\footnote{The following compilers are
available at \texttt{lxplus@cern.ch}: gfortran, lf95 and g95.}.
\item [{\texttt{FORTRAN\_LIBRARIES}}] is the name of the directory in which
your \textsc{Fortran} libraries are located;
in some cases they are automatically found
by the compiler, and you do not need to write them explicitely;
\item [{\texttt{LHAPDFLIB}}] is the absolute path of your Les Houches Accord
PDF Interface library; in a standard installation of LHAPDF, it is
\texttt{/usr/local/lib};
\item [{\texttt{LHAPDFSETS}}] is the absolute path of the directory in which
the PDF sets are stored; in a standard installation of LHAPDF, it is
\texttt{/usr/local/share/lhapdf/PDFsets}.
\end{lyxlist}

In case \textsc{LHAPDF} is not available \textsc{Helac-Phegas} can still run in
standalone mode (see below the definition of the \texttt{lhapdf} keyword) using
the CTEQ6L1 best fit (LO fit with LO $\alpha_s$). In both cases a running
$\alpha_s$ is available (\textsc{LHAPDF} provides its own running $\alpha_s$).

\subsection{How to run a process}

The user interface consists of the following files:

\begin{lyxlist}{00.00.0000}
\item [{\texttt{run.sh}}] is a bash script that reads the input
files, compiles \textsc{Helac-Phegas} and runs it.
\texttt{run\_lsf.sh} and \texttt{run\_lsf2.sh} are its variants to run
with the LSF system\footnote{\texttt{http://batch.web.cern.ch/batch/}}.
\item [{\texttt{default.inp}}] contains a list of keywords, their default
values and a short comment. The user should not modify this file,
but use it just as a reference.
\item [{\texttt{keywords.list}}] is a list of all the available keywords; it
is needed by \texttt{run.sh}.
\item [{\texttt{user.inp}}] is the only file that the user needs to edit.
Here the user can select the process and modify the default values of
many parameters.%
\footnote{The user is free to rename this file. We will refer to
it as \texttt{user.inp} in the following just for clarity.%
}
\item [{\texttt{getqcdscale.h}}] this is the file where the definition of the
QCD scale to be used in the structure function as well as in the
$\alpha_s$ can be set\emph{ by the user}. By default (no editing is
required) it will use $Q_{\rm hard}=M_Z$.
%For a more detailed discussion see section~\ref{matching}.
\item [{\texttt{sub\_lsf}} and {\texttt{sub\_lsf2}}] are the bash scripts that
 send the jobs to the LSF system.
\end{lyxlist} Let us now illustrate the structure of \texttt{user.inp} with an
example.

\begin{verbatim}
# Compulsory information
colpar 1          # colliding particles: 1=pp, 2=ppbar, 3=e+e-
inist  3 -3       # initial state; enter 0 to sum over initial states
finst  11 -11     # final state
energy 14000      # collision energy (GeV)

# For reference, here is the particle numbering:
# ve e u d vm mu c s vt ta t  b  photon z  w+ w- g  h  f+ f- chi jet
# 1  2 3 4  5  6 7 8  9 10 11 12 31     32 33 34 35 41 42 43 44 100
# The respective antiparticles have a minus sign (for example: positron is -2)
# A jet in the final state is denoted by the number 100

# Enter here your additional commands if you wish to alter the default values
\end{verbatim}

With the first four keywords we select the process we want to run
and the collision energy. In the example,
we run $u\overline{u}\rightarrow t\overline{t}$ as part of a
proton-proton collision at $14\,\mbox{TeV}$.

Entering two initial state particles with the \texttt{inist}
keyword, we have implicitly chosen the \emph{single process mode} of
\textsc{Helac-Phegas}. The other possibility is the \emph{summation
mode}, where the initial state particles are hadrons ($pp$ in the
example). \textsc{Helac-Phegas} will find all the partonic processes
that produce the selected final state and will sum over them. To
achieve this, the user should replace the third line in the example
with

\begin{verbatim}
inist  0          # initial state; enter 0 to sum over initial states
\end{verbatim}

In this example, \textsc{Helac-Phegas} finds 9 partonic subprocesses
that produce $t\bar{t}$ in the final state (i.e. the partonic
initial state can be $q\bar{q}$ and $\bar{q}q$ for $q=u,d,s,c$, and
$gg$). The $b$ quark,  by default, is not taken into account as an
initial state parton, but this can be changed with the \texttt{qnum}
keyword which defines the number of quark flavors considered in the
initial and final state, as described in the next subsection.

The summation over partonic subprocesses can also be performed in
the final state. The particle identification ID=$100$ which we call
a 'jet' has the following meaning: the programme will automatically
find all parton level subprocesses, replacing this ID=$100$ with
either a gluon or a quark (anti-quark). Which quark flavors are
taken into account in this subprocesses generation is again defined
by the \texttt{qnum} keyword. For example, if we replace third and
fourth line in the example with

\begin{verbatim}
inist  0          # initial state; enter 0 to sum over initial states
finst  11 -11 100 # final state
\end{verbatim}
\textsc{Helac-Phegas} will look for all partonic subprocesses
having $t\bar{t}+1\,\mbox{parton}$ in the final state. By default,
the $t$ quark is not taken into account as a 'jet' initiator.

In summation mode, the programme calls two auxiliary scripts,
\texttt{script.sh} and \\ \texttt{script.1.sh}, located in the
\texttt{Summation\_Processes} subdirectory.

In case of many jets in the final state, the number of subprocesses
increases rapidly. Since the generation for different subprocesses
is independent, the program is trivially parallelizable. As an
example, we have implemented suitable scripts using the LSF system
at \texttt{lxplus@cern.ch}, see subsection \ref{subsec:LSF} for more
details.

There are many parameters and options that can be changed by the user.
All of them are listed in the \texttt{default.inp} file and described
 in more detail in subsection \ref{subsec:keywords}.

For example,
the line of \texttt{default.inp} referring to the number of Monte
Carlo iterations is

\begin{verbatim}
nmc 100000             # number of montecarlo iterations (single mode)
\end{verbatim}
To change the number of Monte Carlo iterations from 100000 to let say 500000
one has to add the line
\begin{verbatim}
nmc 500000
\end{verbatim}
at the end of the \texttt{user.inp} file. Choices made here will
overwrite the default values.

After editing
the \texttt{user.inp} file, you can compile and run the program by typing\\
{\tt ./run.sh user.inp} \\
or \\
{\tt ./run.sh user.inp myenv-xxx}.\\
In this last case, a file \texttt{myenv-xxx} will be use instead of
\texttt{myenv}.

Summarizing, the user can edit \texttt{user.inp} but is not allowed to
change its structure. In the upper part of the file (lines 1-13) the only
thing the user can do is changing the numerical values of the first four
keywords to select the process. In the lower part of the file (starting
from line 14) the user can
\begin{enumerate}
\item Add lines in the form \texttt{keyword value} to alter the default values;
\item Add empty lines or comment lines.
\end{enumerate}

In case of wrong keywords, repeated keywords or invalid values, the script
will complain and ask the user if she/he wants to continue anyway.

\subsection{Keywords explanation\label{subsec:keywords} }

\subsubsection{Compulsory keywords}

The following keywords must be present in each \texttt{user.inp} file (lines
2-5) in order to select a process which should be run. The symbols,
{\it s, n, b} and {\it x} correspond to the string, integer, boolean (logical)
and real data types respectively.

\begin{itemize}
\item{\noindent \texttt{colpar $n$:}} ~Colliding particles.\\
Set to  $n=1$ for a $pp$ collision, $n=2$ for $p\bar{p}$, $n=3$
for $e^{+}e^{-}$.
\item{\texttt{inist $n_{1}\, n_{2}$:}} ~Particles in the initial state.\\
This keyword implicitly selects the running mode.
If both $n_{1}$ and $n_{2}$ are chosen from Table \ref{tab:Particle-numbering},
\textsc{Helac-Phegas} will run in \emph{single process mode}.
If $n_{1}=0$ and $n_{2}$ is omitted, \textsc{Helac-Phegas}
will run in \emph{summation mode}.
\begin{table}[bh]
\begin{centering}\begin{tabular}{|cc|cc|cc|}
\hline
Number&Particle&Number&Particle&Number&Particle\tabularnewline
\hline
\hline
1&$\nu_{e}$&8&
$s$&
33&
$W^{+}$\tabularnewline
2&
$e^{-}$&
9&
$\nu_{\tau}$&
34&
$W^{-}$\tabularnewline
3&
$u$&
10&
$\tau^{-}$&
35&
$g$\tabularnewline
4&
$d$&
11&
$t$&
41&
$H$\tabularnewline
5&
$\nu_{\mu}$&
12&
$b$&
42&
$\phi_{+}$\tabularnewline
6&
$\mu^{-}$&
31&
$\gamma$&
43&
$\phi_{-}$\tabularnewline
7&
$c$&
32&
$Z$&
44&
$\chi$\tabularnewline
\hline
\end{tabular}\par\end{centering}
\caption{Particle numbering\label{tab:Particle-numbering}}
\end{table}
\item {\texttt{finst $n_{1}\, n_{2}\ldots n_{k}$:}} ~ Particles in the final
state.\\Allowed values are the numbers listed in
Table \ref{tab:Particle-numbering}. In summation mode, $n_{i}=100$
represents a jet.
\item{\texttt{energy $x$:}} ~Collision energy in GeV.
\end{itemize}

\subsubsection{Optional keywords}

The following keywords allow the user to change many default parameters and
options. They can be entered beginning from the 14th line of \texttt{user.inp}.

\paragraph*{General options}

\begin{itemize}

\item{\texttt{exec $s$:}} ~Name of the executable file to be produced.\\
Default name is \texttt{helac-phegas.exe}.
\item{\texttt{outdir $s$:}} ~Name of the output directory.\\ Default name is
\texttt{RESULTS}.
\item{\texttt{nglu} $n$:} ~Number of gluons in the final state.\\ In summation
mode and with jets present in the final state, select only the subprocesses
having $n$ gluons in the final state. Set to $n=-1$ to not constrain the
number of gluons in the final state.\\ Default value is $-1$.
\item{\texttt{qnum} $n_{1}\, n_{2}$:} ~Number of quark flavors in the initial
  and final state.\\ In summation mode, when
looking for the allowed subprocesses \textsc{Helac-Phegas} will take
into account only the $n_{1}$ lightest flavors in the initial state,
and the $n_{2}$ lightest flavors as jet initiators in the final
state.\\ Default values are $n_{1}=4$ ($u,d,c,s$) and $n_{2}=5$
($u,d,c,s,b$).
\item{\texttt{ktrw} $b$:} ~Switch for the $k_\perp$ reweighting algorithm.\\
%The algorithm used in the so-called MLM merging procedure,
See section \ref{matching} for more details. Set
F, f or 0 to switch off the algorithm, or set T, t or 1 to switch it on.\\
Default value is false.
\item{\texttt{lrgnc} $b$:} ~Switch for large $N_c$ limit.\\
 In the limit $N_c\to \infty$, only the diagonal terms of the color
matrix ${\cal {C}}_{IJ}$, cf. Eq. (\ref{colormatrix}), survive,
and all $ {\cal O}(N_{c}^{-2})$ terms
are neglected. The interferences between different colour flows vanish
which speeds up the calculation. This option can be used e.g. to test,
whether the large $N_c$ limit approximation can reliably describe the physical
process\footnote{Let us stress here, that default calculation in
  \textsc{Helac-Phegas}
takes into account the full color structure, no approximation is
  involved.}. Moreover, in the case when \texttt{ktrw}
is switched on, the user still has the option to switch on or off the
large $N_c$ limit.\\ Default value is false.
%\item {\texttt{histo} $s$:} ~Name of the histogram file.\\ Default is
%\texttt{hi\_file}
\item{\texttt{error} $s$:} ~Name of the file, where,
every 1000 iterations, the
  cross section and its Monte Carlo error are recorded. \\ Default is
\texttt{err\_file}.
\item{\texttt{gener} $n$:} Choice of the phase space generator.\\
Besides the default \textsc{Phegas} \cite{Papadopoulos:2000tt} described in
Section~\ref{phegas}, there are two other possibilities. The first one is
\textsc{Rambo} \cite{Kleiss:1985gy}, a flat
phase-space generator. It generates the momenta distributed
uniformly in phase space so that a large number of events is needed to
integrate the integrands to acceptable precision, which results in a rather
low computational efficiency. The second one is \textsc{Durham},
 an algorithm that for a
given order of the final state particles expresses the phase
space in terms of the
$k_{T}$, $\phi$ and $\bar{y}$ variables, where $\bar{y}$ is the difference of
the rapidities of two nearest particles.
Adaptation over all possible orderings of the
final state particles is also included. For more details see
Ref. \cite{Papadopoulos:2005ky}.
Set to $n=0$
to use \textsc{Phegas}, $n=1$ for \textsc{Rambo} or $n=2$ for \textsc{Durham}.
\\ Default value is 0.
\item{\texttt{repeat} $n$:} ~\textsc{Helac-Phegas}
run option.\\ Set to  $n=0$ to run
both phases of \textsc{Helac-Phegas},
$n=1$ to run only the first phase, where the
skeleton for a given process is constructed, or $n=2$ to
run only the second phase. \\ Default value is 0.
\item{\texttt{ranhel} $n$:} ~Choice of the set-up for the helicity
configurations. \\ There are two possibilities, either exact
summation over all helicity configurations is performed or Monte
Carlo integration is applied~\cite{Kanaki:2000ey}.  For example for
a massive gauge boson the second option is achieved by introducing
the polarization vector
\begin{equation}
\varepsilon^{\mu}_{\phi}(p)=e^{i\phi}\varepsilon^{\mu}_{+}(p)+
e^{-i\phi}\varepsilon^{\mu}_{-}(p)+\varepsilon^{\mu}_{0}(p),
\end{equation}
where $\phi \in (0,2\pi)$ is a random number. By integrating over
$\phi$ the correct sum over helicities is obtained
\[
\frac{1}{2\pi}\int_{0}^{2\pi}d\phi ~\varepsilon^{\mu}_{\phi}(p)
(\varepsilon^{\nu}_{\phi}(p))^{*}=\sum_{\lambda=\pm}
\varepsilon_{\lambda}^{\mu}(p)(\varepsilon_{\lambda}^{\nu}(p))^{*}.
\]
The same idea is applied to the helicity of (anti)fermions. Set to
$n=0$ to calculate explicitly all  helicity configurations or $n=1$
to do a Monte Carlo over helicities.\\ Default value is 1.
\item{\texttt{alphasrun} $n$:} ~Switch for the running of $\alpha_{s}$.\\
Set to $n=0$ to have a fixed value of $\alpha_{s}$ or $n=1$ to have
a running coupling $\alpha_{s}(Q^{2})$. In the later case the $Q$ scale has
to be implemented by the user in the {\tt getqcdscale.h} file, where a
few examples have already been provided.\\
Default value is 0.
\item{\texttt{gauge} $n$:} ~Choice of the gauge.\\
Set to $n=0$ for the Feynman gauge or $n=1$ for the unitary gauge.\\
Default value is 1.
\item{\texttt{ihiggs} $b$:} ~Switch for the
inclusion of the contribution coming  from the Higgs boson.\\
Default value is false.
\item{\texttt{widsch} $n$:} ~Scheme for the introduction of the width of $W$,
$Z$ bosons as well as the top
quark\cite{Argyres:1995ym,Beenakker:1996kn,Denner:2006ic}. Set to $n=0$
for the fixed width scheme or $n=1$ for the complex  mass scheme.\\
Default value is 0.
\item{\texttt{qcd} $n$:} ~Option for the type of  interactions.\\
Set to $n=0$ to have only
electroweak interactions, $n=1$ to have both electroweak and QCD and  $n=2$
for QCD only.\\ Default value is 1.
\item{\texttt{unwgt} $b$:} ~Switch for the un-weighting procedure.\\ If set to
false events are not unweighted. If set to true events are unweighted.\\
Default value is true.
\item{\texttt{preunw} $n$:}  ~Pre-unweighted events.\\ Number of events
generated to calculate the maximal weight to start the un-weighting
procedure. It is recommended that the user tries different values in order
to obtain a better un-weighting efficiency. \\ Default value is 10000.
\item{\texttt{unwevt} $n$:} ~Number of unweighted events to be generated.\\ In
the summation mode, $n$ is the total number of unweighted events, which is
distributed between all possible parton level processes according to their
relative contribution to the total cross section.\\ Default value is 50000.
\item{\texttt{nmc} $n$:} ~Number of Monte Carlo iterations in the
single process mode.\\ Not used in summation mode.\\ Default value is 100000.
\item{\texttt{nmc1} $n$:} ~Number of Monte Carlo iterations in the first step
  of the summation mode. Not used in single process mode.\\
Default value is 100000.
\item{\texttt{nmc2} $n$:} ~Number of Monte Carlo iterations in the second step
  of the summation mode. Not used in single process mode or if un-weighting is
switched off.\\
Default value is 3000000.
\item{\texttt{optim} $n_{1}\, n_{2}\, n_{3}\, n_{4}\, n_{5}\, n_{6}$:}
~Optimization options for the \textsc{Phegas} phase space generator.
Optimization is performed when the program reaches
the Monte Carlo iterations\[ \left\{
o_{1}=n_{1},o_{2}=o_{1}+n_{2}n_{3},o_{3}=o_{2}+n_{2}n_{3}^2,
\ldots,o_{n_{4}}=o_{n_{4}-1}+n_{2}n_{3}^{n_4-1}\right\} .\] It stops when the
maximum
number of optimizations $n_{4}$ is reached, or when the $n_{5}$-th Monte Carlo
iteration is reached. $n_{6}$ is a flag: if set to 0 optimisation is not
performed, if set to 1 it is performed.\\ Default values are 10000 10000 1 8
100000 1.
\item{\texttt{lhapdf} $b$:} ~Switch for the use of the Les Houches Accord PDF
interface library.\\ If set to F, \textsc{Helac-Phegas}
will run in standalone mode, using the CTEQ6L1 best fit
(LO fit with LO $\alpha_s$).\\ Default value is false.
\item{\texttt{pdf} $n$:} ~Choice of the PDF set.\\ Used only if
\texttt{lhapdf} is set to true. $n$ must be equal to the  LHAGLUE number
of the desired PDF set (see PDFsets.index or Ref.~\cite{Whalley:2005nh}).
Enter $n=0$ to run without PDFs.\\ Default value is 10042, which corresponds
to CTEQ6L1 (LO fit with LO $\alpha_s$).
\item{\texttt{pythia} $n_{1}\, n_{2}$:} {\tt IDWTUP} and {\tt NPRUP}
parameters for the LHA sample files.\\ {\tt IDWTUP} dictates how the event
weights are interpreted by the showering and hadronization event generators.
{\tt NPRUP} is the number of different user processes, see \cite{Boos:2001cv}
for more details.\\
Default values are $n_{1}=3$, which corresponds to an option with
unweighted events with the weight = +1, and $n_{2}=1$, which corresponds to
the one process only.
\item{\texttt{constants} $b$:} if $b=1$ the user can provide its own
\texttt{constants.h} file in accordance to the prototype given in the
\texttt{constants\_std.h} file.\\
Default value is $b=0$, see below the description of physical constants.
\end{itemize}

\paragraph*{pp(bar) cuts\protect \footnote{Cuts not already provided may be
added in the {\tt 2.cuts\_auto\_pp.f} file.} \\ }
The following options
%in Table \ref{tab:ppcuts}
are read only if \texttt{colpar} is set to 1 or 2
(respectively, $pp$ and $p\bar{p}$). All dimensional quantities are in GeV.
When the word 'quark' appears, light quarks and gluons have to be
understood.
%\begin{table}[ht]
\begin{center}\begin{tabular}{|c|c|c|}

\hline keyword& description& default\tabularnewline \hline \hline

\texttt{\textbf{cutoffp}} $x$& Cutoff value& 1.0d-3\tabularnewline
\texttt{\textbf{minptl}} $x$& Minimum lepton $p_{T}$& 1.0d-3\tabularnewline
\texttt{\textbf{minptq}} $x$& Minimum quark $p_{T}$& 8.0\tabularnewline
\texttt{\textbf{minptb}} $x$& Minimum bottom $p_{T}$& 8.0\tabularnewline
\texttt{\textbf{minptt}} $x$& Minimum top $p_{T}$& 0.0\tabularnewline
\texttt{\textbf{minptp}} $x$& Minimum photon $p_{T}$& 8.0\tabularnewline
\texttt{\textbf{maxrapl}} $x$& Maximum lepton rapidity& 10.0\tabularnewline
\texttt{\textbf{maxrapq}} $x$& Maximum quark rapidity& 2.0\tabularnewline
\texttt{\textbf{maxrapb}} $x$& Maximum bottom rapidity& 2.0\tabularnewline
\texttt{\textbf{maxrapt}} $x$& Maximum top rapidity& 10.0\tabularnewline
\texttt{\textbf{maxrapp}} $x$& Maximum photon rapidity& 2.0\tabularnewline
\texttt{\textbf{mindrll}} $x$& Minimum $\Delta R$ separation between lepton
and lepton& 0.0\tabularnewline
\texttt{\textbf{mindrlq}} $x$& Minimum $\Delta R$ separation between lepton
and quark& 0.0\tabularnewline
\texttt{\textbf{mindrqq}} $x$& Minimum $\Delta R$ separation between quark and
quark& 0.7\tabularnewline
\texttt{\textbf{mindrqb}} $x$& Minimum $\Delta R$ separation between quark and
bottom& 0.7\tabularnewline
\texttt{\textbf{mindrbb}} $x$& Minimum $\Delta R$ separation between bottom
and bottom& 0.7\tabularnewline
\texttt{\textbf{mindrpf}} $x$& Minimum $\Delta R$ separation between photon
and fermion& 0.0\tabularnewline
\texttt{\textbf{minmqqp}} $x$& Minimum quark-quark invariant mass in the
$pp/p\bar{p}$ case& 0.0\tabularnewline
\texttt{\textbf{minmqb}} $x$& Minimum quark-bottom invariant mass&
0.0\tabularnewline
\texttt{\textbf{minmbb}} $x$& Minimum bottom-bottom invariant mass&
0.0\tabularnewline
\hline
\end{tabular}\par\end{center}

\paragraph*{$ \mathbf {e^{+}e^{-}}$ cuts\protect \footnote{Cuts not already
provided may be added in the {\tt 2.cuts\_auto\_ee.f} file.}\\ }
The following options
%in Table \ref{tab:eecuts}
are read only if \texttt{colpar} is set to 3
($e^{+}e^{-}$). All dimensional quantities are in GeV.

%\begin{table}[ht]
\begin{center}\begin{tabular}{|c|c|c|}
\hline keyword& description& default\tabularnewline \hline \hline
\texttt{\textbf{cutoffe}} $x$& Cutoff value& 1.0d-3\tabularnewline
\texttt{\textbf{minenl}} $x$& Minimum lepton energy& 10.0\tabularnewline
\texttt{\textbf{minenq}} $x$& Minimum quark energy& 10.0\tabularnewline
\texttt{\textbf{minenp}} $x$& Minimum photon energy& 10.0\tabularnewline
\texttt{\textbf{minanglb}} $x$& Minimum angle (degrees) between lepton and
beam& 5.0\tabularnewline
\texttt{\textbf{minangqb}} $x$& Minimum angle (degrees) between quark and
beam& 5.0\tabularnewline
\texttt{\textbf{minangpb}} $x$& Minimum angle (degrees) between photon and
beam& 5.0\tabularnewline
\texttt{\textbf{minangll}} $x$& Minimum angle (degrees) between lepton and
lepton& 5.0\tabularnewline
\texttt{\textbf{minanglq}} $x$& Minimum angle (degrees) between lepton and
quark& 5.0\tabularnewline
\texttt{\textbf{minangqq}} $x$& Minimum angle (degrees) between quark and
quark& 5.0\tabularnewline
\texttt{\textbf{minangpf}} $x$& Minimum angle (degrees) between photon and
fermion& 5.0\tabularnewline
\texttt{\textbf{minmqq}} $x$& Minimum quark-quark invariant mass in the
$e^{+}e^{-}$ case& 10.0\tabularnewline

\hline

\end{tabular}\par\end{center}

\paragraph*{Physical constants\protect \\ }

The user can set the values of many physical constants using the keywords
 reported in Table \ref{tab:constants}.

By default the electroweak couplings are given by
\[
    sin^2\theta_w=1-\left( \frac{m_W^2}{m_Z^2}\right) ,\,\,
    g_{weak}=(4\sqrt{2} G_F)^{1/2} m_W,\,\,\,
    \alpha_{em}=\sqrt{2}G_F m_W^2 sin^2\theta_w / \pi
\]

However, the user can optionally give values for $\alpha_{em}$
(\texttt{alphaem}) and $sin^2\theta_w$ (\texttt{sin2thetaw}).

For more advanced users, it is recommended to edit and provide their
 own \texttt{constants.h} file. A prototype is
included, see the  \texttt{constants\_std.h} file. The
CKM~\cite{Cabibbo:1963yz,Kobayashi:1973fv}
matrix is also possible to be defined in this file, as well as masses
and widths for all particles.

\begin{table}[ht]
\begin{center}\begin{tabular}{|c|c|c|}
\hline keyword& description& default\tabularnewline \hline \hline
\texttt{\textbf{gfermi    }} $x$& Fermi coupling constant&
1.16639d-5\tabularnewline
\texttt{\textbf{sin2thetaw}} $x$& Sinus squared of Weinberg angle
$sin^2\theta_w$&
$-1^{*}$\tabularnewline
\texttt{\textbf{alphaem}} $x$& Electromagnetic coupling constant&
$-1^{*}$\tabularnewline
\texttt{\textbf{alphas2}} $x$& Fixed strong coupling constant, used only if
\texttt{alphasrun} is 0& 0.118\tabularnewline
\texttt{\textbf{zmass}} $x$& $Z$ mass& 91.188\tabularnewline
\texttt{\textbf{zwidth}} $x$& $Z$ width& 2.446\tabularnewline
\texttt{\textbf{wmass}} $x$& $W$ mass& 80.419\tabularnewline
\texttt{\textbf{wwidth}} $x$& $W$ width& 2.048\tabularnewline
\texttt{\textbf{higmass}} $x$& Higgs boson mass& 130.0\tabularnewline
\texttt{\textbf{higwidth}} $x$& Higgs boson width& 4.291d-3\tabularnewline
\texttt{\textbf{emass}} $x$& $e$ mass& 0.0\tabularnewline
\texttt{\textbf{mumass}} $x$& $\mu$ mass& 0.0\tabularnewline
\texttt{\textbf{taumass}} $x$& $\tau$ mass& 0.0\tabularnewline
\texttt{\textbf{umass}} $x$& $u$ quark mass& 0.0\tabularnewline
\texttt{\textbf{dmass}} $x$& $d$ quark mass& 0.0\tabularnewline
\texttt{\textbf{smass}} $x$& $s$ quark mass& 0.0\tabularnewline
\texttt{\textbf{cmass}} $x$& $c$ quark mass& 0.0\tabularnewline
\texttt{\textbf{bmass}} $x$& $b$ quark mass& 0.0\tabularnewline
\texttt{\textbf{tmass}} $x$& $t$ quark mass& 174.3\tabularnewline
\texttt{\textbf{twidth}} $x$& $t$ quark width& 1.6\tabularnewline

\hline

\end{tabular}\par\end{center}
\caption{Keywords and default values regarding some physical constants.
\newline $^*$The value $-1$ for the keywords \texttt{sin2thetaw} and
\texttt{alphaem} means that they are defined as described in the text. }
\label{tab:constants}
\end{table}

\subsection{How to run HELAC  using the LSF system}
\label{subsec:LSF}

Running \textsc{Helac-Phegas} on
a single machine can become impractical when the
final state contains  many particles and jets, either because of time
constraints or because of the amount of output that can be produced.
A possible solution to this problem is to run the $n$ involved subprocesses at
the same time on a computer cluster. We provide scripts to run
\textsc{Helac-Phegas} on the CERN's batch system
%\footnote{Let us stress here,
%that an account at CERN is required in this case.}.

The keywords \texttt{queue1} and \texttt{queue2} allow the user to choose
the queue for the
subprocesses in the first and second phase of the \textsc{Helac-Phegas} run.

By default, jobs are sent to the \texttt{8nm} queue, 8 minutes, which
is suitable only for short runs. Two options are available:
\begin{enumerate}

\item Write everything, also the intermediate files, which can be quite large,
in the AFS area.

In this case, a \texttt{user.inp} file has to be prepared as usual, and then
\begin{verbatim}
./sub_lsf
\end{verbatim}
should be run.
The user will be asked to provide some information like
job name, queue, user input file and environment file.
The queue will only be used for
the controller job, and may, therefore, be short.

\item Write everything on the remote, \texttt{lxb}, hosts (suggested method).

At the end of the run, the sample will be written in the AFS directory from
which the job has been submitted. The usual output directory will be written,
in compressed form, in the user's personal CASTOR area.

Similarly to the previous case, a \texttt{user.inp} file has to be prepared
and then
\begin{verbatim}
./sub_lsf2
\end{verbatim}
should be run.
The user has to provide a job name, queue, user input file and
environment file.
The script will automatically resubmit a job to a longer queue, if it
failed. Notice, however, that this is only useful in the case of a failure
due to time constraints.
\end{enumerate}

There are two reasons for using this method:
\begin{enumerate}
\item To save time. More subprocesses run at the same time, while with the
standard procedure they run one after the other.
\item To save space. The output files produced can exceed your disk quota.
\end{enumerate}

The first procedure (\texttt{sub\_lsf}) saves time but not space, whereas the
second one (\texttt{sub\_lsf2}) saves also space. Another reason to prefer the
second procedure is the automatic resubmission of failed jobs, which only is
implemented in this case.

\subsection{Description of the output}

All the outputs produced by the \textsc{Helac-Phegas} program
 will be written in a subdirectory, whose name
is defined by the keyword \texttt{outdir}. Let us distinguish two main cases.

\subsubsection{Single process mode}
In the output directory you will find the following files.
\begin{itemize}
\item{Executable files.} The \textsc{Helac-Phegas} executable (whose name is
defined by the keyword \texttt{exec}) and, only if \texttt{unwgt} is true, the
executable for the un-weighting procedure, \texttt{unwei.exe}.
\item User input file, called \texttt{user.inp} throughout this paper.
\item{\texttt{input\_sp} and \texttt{output}.} Respectively input and output
redirected files of the \textsc{Helac-Phegas} executable. If \texttt{unwgt} is
true, the output of \texttt{unwei.exe} is appended to \texttt{output}.
\item{The error file,} whose name is defined by the keyword
\texttt{error}. Every 1000 iterations, the cross section and its error are
recorded to this file. The three columns are respectively the number of
iterations, the cross section and the error.

\item{The kinematic file,} \texttt{kine\_XXX.out}, where \texttt{XXX}
represents the process. Used for debugging.
\item{The event sample file,} \texttt{sampleXXX.lhe}, where \texttt{XXX}
represents the process, in the standard format dictated by the Les Houches
accord \cite{Boos:2001cv,Alwall:2006yp}. It is generated only if
\texttt{unwgt} is set to true.
\end{itemize}
\subsubsection{Summation mode}
In the output directory you will find
\begin{itemize}
\item{The \textsc{Helac-Phegas} executable.}
\item{User input file.}
\item{\texttt{infile},} the input file for \texttt{processes.f}, that
generates all the subprocesses.
\item{The (gzipped) sample file,} \texttt{sampleXXX.lhe.gz}, where
\texttt{XXX} describes the final state. It is generated only
if \texttt{unwgt} is set to true.
\item{A subdirectory \texttt{1},} where all the outputs generated in the
first phase, generation of weighted events, will be written.
\item{A subdirectory \texttt{2},} where all the outputs generated in the
second phase, generation of unweighted events, will be written. It is
generated only if \texttt{unwgt} is set to true.
\end{itemize}
The subdirectory \texttt{1} contains
\begin{itemize}
\item{\texttt{input},} a skeleton of input files, needed to generate the real
input files for each single subprocess.
\item{\texttt{output},} the output of \texttt{readoutput.f}.
\item{\texttt{results.output}.} Each group of two lines refers
to a single subprocess. The first
line is the cross section and the second is the error.
\item{\texttt{cross\_sections}.} The $n$-th line shows
the percentage of the total cross section
represented by the $n$-th subprocess.
\item{Subdirectories \texttt{directory\_n},} one for
each subprocess. The content of each of these
directories resembles the one described for the single
process mode, except that most of the files
are gzipped to save space.
\end{itemize}

The content of the subdirectory \texttt{2} is analogous
to \texttt{1}, with the main exception that
in the \texttt{results.output} file a third and a fourth line for
each subprocesses is added, showing information on generated unweighted events.

\subsubsection{Output of the main \textsc{Helac-Phegas} executable}

A substantial part of the information contained in the output of
{\tt helac-phegas.exe} is self-explainable. In this section we would like to
make the user familiar with the main points of this file.
Lets take as the basic example the following process
$u(p_1)\bar{u}(p_2) \to g(p_3) g(p_4)$.

The output file begins with the reproduction of the input file, followed by
masses and widths of all SM particles. Then we have
\begin{verbatim}
 QCD INCLUDED, g= 1.217715784776720 0.1180000000000000
 UqUaG0G0
 TIME= 2007 10 8 180 19 26 52 966
 avhel,avcol,symet
 0.2500000000000000 0.1111111111111111 0.5000000000000000
\end{verbatim}
where the process is clearly stated {\tt UqUaG0G0}, together with the QCD
coupling and factors that correspond to the average over
initial helicities, colors and the symmetry factor.
Then the 'first phase' starts. For each color connection configuration
the color connection assignments for all particles is given, namely
\begin{verbatim}
 the colour of particles ONE 2 0 3 1
 the colour of particles TWO 0 1 2 3
\end{verbatim}
the first line is the color connection the second the anti-color one (kept
fixed at the ordinal permutation).
Then the solution for the Dyson-Schwinger equations is presented in the
following  form
\begin{verbatim}
for the     3     2  colour conf. there are       8  subamplitudes
   1   2   6  -3   5   1   1   4  35   3   2  -3   2   0   0   0   1   1   1
   2   2   6  -3   5   0   1   4  35   3   2  -3   2   0   0   0   2   1   1
   3   2  10  -3   6   1   1   8  35   4   2  -3   2   0   0   0   1   1   3
   4   2  10  -3   6   0   1   8  35   4   2  -3   2   0   0   0   2   1   3
   5   2  14  -3   7   1   2   4  35   3  10  -3   6   0   0   0   1   1   1
   6   2  14  -3   7   0   2   4  35   3  10  -3   6   0   0   0   2   1   1
   7   2  14  -3   7   2   2   8  35   4   6  -3   5   0   0   0   1   1   3
   8   2  14  -3   7   0   2   8  35   4   6  -3   5   0   0   0   2   1   3
\end{verbatim}
Let us focus on the line number one
\begin{verbatim}
   1   2   6  -3   5   1   1   4  35   3   2  -3   2   0   0   0   1   1   1
\end{verbatim}
It simply encodes all information -- to be used internally -- corresponding to
the fusion of an $\bar{u}$ with momentum $p_2$ with the $g$ with momentum
$p_3$ to produce an $\bar{u}$ with total momentum $p_2+p_3$ (all momenta
taken incoming). At the end of this part a line with
\begin{verbatim}
 the number of Feynman graphs =  3
 the number of Feynman graphs =  3
\end{verbatim}
followed by the color matrix is given. The number of Feynman graphs for this
process is of course three. Then the number of channels to be used by
\textsc{ Phegas} (always the number of graphs +1) is given with the number
of MC points {\tt nmc} and optimization parameters.

 \begin{verbatim}
 Number of channels 4
 Number of MC points
  nmc  =  100000
 nopt,nopt_step,optf,maxopt,iopt
 100 10000 1.000000000000000 8 1
 number of channels= 4
 NUMBER OF CHANNELS 4
 \end{verbatim}
Since \textsc{Phegas} is using Feynman graphs as the basic structure of the
multichannel phase-space mappings, a description of each Feynman graph is
given as follows:
\begin{verbatim}
 the graph 1
 14 -3 12 35 2 -3 0 0
 12 35 4 35 8 35 0 0
\end{verbatim}

After a few lines with self-explained information the cuts
\begin{verbatim}
 ---------------------------------------------------
         the cuts
 pt     of   3    particle    8.000000000000000
 energy of   3    particle    8.000000000000000
 pt     of   4    particle    8.000000000000000
 energy of   4    particle    8.000000000000000
 rapidity of   3    particle    2.000000000000000
 rapidity of   4    particle    2.000000000000000
 cos-beam1 of  3    particle    0.9640275800758169
 cos-beam1 of  4    particle    0.9640275800758169
 cos-beam2 of  3    particle    0.9640275800758169
 cos-beam2 of  4    particle    0.9640275800758169
 DR      3   with   4 0.7000000000000000
 cos of  3   with   4 0.7648421872844885
 mass of  3   with   4 5.486364919287221
 ---------------------------------------------------
\end{verbatim}
applied are given for each particle or pair of particles. Then a bunch of
lines follows in the form

\begin{verbatim}
sigma=   0.175490D+03   0.709135D-01       491      1000      1000
 ----------------------------
\end{verbatim}
where the calculated cross section {\tt sigma} with the percentage error,
the number of events passing the cuts, the number of phase-space points used
as well as the number of phase-space points tried, is given.

At the end of the file one normally has
\begin{verbatim}
 out of  100000    100001  points have been used
 and   50419  points resulted to =/= 0 weight
 whereas   49582  points to 0 weight
  estimator x:    0.182757D+03
  estimator y:    0.147435D+01
  estimator z:    0.124416D-02
  average estimate :   0.182757D+03
                +\-    0.121423D+01
  variance estimate:   0.147435D+01
                +\-    0.352726D-01
 lwri: points have used 0.000000000000000E+00
 2212 2212 7000.000000000000 7000.000000000000 3 1
 % error: 0.6643944554436630
\end{verbatim}

which is translated to the total cross section in nb \texttt{0.182757D+03}
the MC error also in nb \texttt{0.121423D+01} and the percent
error \texttt{0.6643944554436630}\%.

If the \texttt{unwgt} is set to true the distribution of the weights
is plotted for the  number of events given by \texttt{preunw}. Then the
un-weighting procedure starts and the number of unweighted events is
additionally printed. In case of un-weighting the final histogram of
 the weights is also printed.

\subsection{External phase space points}

In this mode the user can provide by herself/himself momenta for the particles
 participating in the process under consideration and get back information
on the squared matrix element summed over spin and color degrees of freedom.
Additionally, the color ordered amplitude for a given color connection and
helicity configuration can be obtained.

\begin{itemize}

\item{\texttt{onep:}} ~Set to true in order to have the described option.
\\ Default value is false.
\item{\texttt{mom:}} ~Name of the input file provided by the user with the
momenta of the particles, in the format $p_x,p_y,p_z,E,m,w$, where $w$ is a
weight for phase space generation, usually set $w=1$. A prototype is included
under the name \texttt{mom}.
    \\ Default name is \texttt{mom}.
\item{\texttt{momout}:} ~Name of the output file.\\ Default name is
\texttt{momout}.

\end{itemize}

\subsection{Benchmarks}

From the physics point of view \textsc{Helac-Phegas} has already been used
in several contexts to produce results in physics. A few examples are: the
Monte Carlo generator \textsc{Nextcalibur}~\cite{Berends:2000fj} that
includes the  \textsc{Excalibur}~\cite{Berends:1994xn} phase space generation
and \textsc{Helac-Phegas} matrix elements,
used in $e^+e^-$ analysis; the extensive
comparison with \textsc{Sherpa} on six-fermion production
processes~\cite{Gleisberg:2003bi}; and more recently the participation in
the $W+n$ jets comparative study \cite{Alwall:2007fs} of different Monte
 Carlo codes namely
\textsc{Alpgen}, \textsc{Ariadne}, \textsc{Helac-Phegas},
\textsc{MadEvent} and \textsc{Sherpa}.

On the technical side, we provide a number of benchmark calculations in the
web-page of the code, so that the potential user can test and validate his
own results.

\section{Outlook}

The current version is named {\tt 1.2.0}; it is meant to be publicly
used: in that sense we welcome any bug report or simple question.
The first number introduces major changes, the first one described
below. The second number refers to minor changes and/or additions,
whereas the last one to bug corrections.

A version with all Higgs-gluon and Higgs-photon couplings, in the
large $m_{top}$ limit will follow as {\tt 1.3.0}. It is already
tested and will be available soon on the \textsc{Helac-Phegas}
web-page.

A new version suitable for processes with many colored particles, i.e.
 with a number of
equivalent gluons more than 9, for instance $gg\to 8g$ and more, has been
already developed and tested~\cite{Papadopoulos:2005ky}. It will be
incorporated in version 2.

We are also working on the inclusion of MSSM particles and couplings.

\section*{Acknowledgments}
We would like to thank Andre van Hameren for providing us with his
\textsc{Parni} routines and Michelangelo Mangano for useful discussions on
matching algorithms. We would like to acknowledge support from the Transfer of
Knowledge programme {\tt ALGOTOOLS} (MTKD-CT-2004-014319) and from the RTN
{\tt HEPTools} (MRTN-2006-CT-035505). M.W.
was supported in part
by BMBF grant  05 HT6VKC. She also would like to thank the Galileo Galilei
Institute for Theoretical Physics for its hospitality and the INFN for partial
support during the completion of this work.

%\bibliographystyle{h-elsevier3}
%\bibliography{helac}

\begin{thebibliography}{10}

\bibitem{Berends:1987me}
F. A. Berends and W. T. Giele,
\newblock Nucl. Phys. B306 (1988) 759.
%%CITATION = NUPHA,B306,759;%%

\bibitem{Berends:1988yn}
F. A. Berends, W. T. Giele and H. Kuijf,
\newblock Nucl. Phys. B321 (1989) 39.
%%CITATION = NUPHA,B321,39;%%

\bibitem{Berends:1990ax}
F. A. Berends, H. Kuijf, B. Tausk and W. T. Giele,
\newblock Nucl. Phys. B357 (1991) 32.
%%CITATION = NUPHA,B357,32;%%

\bibitem{Argyres:1992js}
E. N. Argyres, R. H. P. Kleiss and C. G. Papadopoulos,
\newblock Nucl. Phys.  B391 (1993) 42.
%%CITATION = NUPHA,B391,42;%%

\bibitem{Argyres:1993wz}
  E. N. Argyres, R. H. P. Kleiss and C. G. Papadopoulos,
\newblock  Phys. Lett.  B308 (1993) 292
  [Addendum-ibid.  B319 (1993) 544], hep-ph/9303321.
  %%CITATION = PHLTA,B308,292;%%

\bibitem{Caravaglios:1995cd}
F. Caravaglios and M. Moretti,
\newblock Phys. Lett. B358 (1995) 332, hep-ph/9507237.
%%CITATION = HEP-PH/9507237;%%

\bibitem{Draggiotis:1998gr}
P. Draggiotis, R. H. P. Kleiss and C. G. Papadopoulos,
\newblock Phys. Lett. B439 (1998) 157, hep-ph/9807207.
%%CITATION = HEP-PH/9807207;%%

\bibitem{Caravaglios:1998yr}
F. Caravaglios, M. L. Mangano, M. Moretti and R. Pittau,
\newblock Nucl. Phys. B539 (1999) 215, hep-ph/9807570.
%%CITATION = NUPHA,B539,215;%%

\bibitem{Kanaki:2000ey}
A. Kanaki and C.G. Papadopoulos,
\newblock Comput. Phys. Commun. 132 (2000) 306, hep-ph/0002082.
%%CITATION = HEP-PH/0002082;%%

\bibitem{Kanaki:2000ms}
A. Kanaki and C. G. Papadopoulos,
\newblock hep-ph/0012004.
%%CITATION = HEP-PH/0012004;%%

\bibitem{Draggiotis:2002hm}
P. D. Draggiotis, R. H. P. Kleiss and C. G. Papadopoulos,
\newblock Eur. Phys. J. C24 (2002) 447, hep-ph/0202201.
%%CITATION = HEP-PH/0202201;%%

\bibitem{Papadopoulos:2005ky}
C. G. Papadopoulos and M. Worek,
\newblock Eur. Phys. J. C50 (2007) 843, hep-ph/0512150.
%%CITATION = HEP-PH/0512150;%%

\bibitem{Draggiotis:2006er}
P. Draggiotis, A. van Hameren, R. Kleiss, A. Lazopoulos, C. G. Papadopoulos
and M. Worek,
\newblock Nucl. Phys. Proc. Suppl. 160 (2006) 255, hep-ph/0607034.
%%CITATION = NUPHZ,160,255;%%

\bibitem{Duhr:2006iq}
C. Duhr, S. Hoche and F. Maltoni,
\newblock JHEP 0608 (2006) 062, hep-ph/0607057.
%%CITATION = JHEPA,0608,062;%%

\bibitem{Papadopoulos:2000tt}
C. G. Papadopoulos,
\newblock Comput. Phys. Commun. 137 (2001) 247, hep-ph/0007335.
%%CITATION = HEP-PH/0007335;%%

\bibitem{Kleiss:1994qy}
  R. Kleiss and R. Pittau,
\newblock  Comput. Phys. Commun. 83 (1994) 141, hep-ph/9405257.
  %%CITATION = CPHCB,83,141;%%

\bibitem{Mangano:2002ea}
M. L. Mangano, M. Moretti, F. Piccinini, R. Pittau and A. D. Polosa,
\newblock JHEP 0307 (2003) 001, hep-ph/0206293.
%%CITATION = JHEPA,0307,001;%%

\bibitem{Mangano:2001xp}
M. L. Mangano, M. Moretti and R. Pittau,
\newblock Nucl. Phys. B632 (2002) 343, hep-ph/0108069.
%%CITATION = HEP-PH/0108069;%%

\bibitem{Lonnblad:1992tz}
L. Lonnblad,
\newblock Comput. Phys. Commun. 71 (1992) 15.
%%CITATION = CPHCB,71,15;%%

\bibitem{Stelzer:1994ta}
T. Stelzer and W. F. Long,
\newblock Comput. Phys. Commun. 81 (1994) 357, hep-ph/9401258.
%%CITATION = HEP-PH/9401258;%%

\bibitem{Maltoni:2002qb}
F. Maltoni and T. Stelzer,
\newblock JHEP 0302 (2003) 027, hep-ph/0208156.
%%CITATION = HEP-PH/0208156;%%

\bibitem{Alwall:2007st}
J. Alwall et al.,
\newblock  JHEP 0709 (2007) 028, arXiv:0706.2334 [hep-ph].
%%CITATION = JHEPA,0709,028;%%

\bibitem{Krauss:2001iv}
F. Krauss, R. Kuhn and G. Soff,
\newblock JHEP 0202 (2002) 044, hep-ph/0109036.
%%CITATION = HEP-PH/0109036;%%

\bibitem{Gleisberg:2003xi}
T. Gleisberg, S. Hoche, F. Krauss, A. Schalicke,
S. Schumann and J. C. Winter,
\newblock JHEP 0402 (2004) 056, hep-ph/0311263.
%%CITATION = HEP-PH/0311263;%%

\bibitem{Kilian:2007gr}
W. Kilian, T. Ohl and J. Reuter,
\newblock  arXiv:0708.4233 [hep-ph].
%%CITATION = ARXIV:0708.4233;%%

\bibitem{Pumplin:2002vw}
J. Pumplin, D. R. Stump, J. Huston, H. L. Lai, P. M. Nadolsky and W. K. Tung,
\newblock  JHEP 0207 (2002) 012, hep-ph/0201195.
%%CITATION = JHEPA,0207,012;%%

\bibitem{Stump:2003yu}
D. Stump, J. Huston, J. Pumplin, W. K. Tung, H. L. Lai, S. Kuhlmann and J. F. Owens,
\newblock  JHEP 0310 (2003) 046, hep-ph/0303013.
%%CITATION = JHEPA,0310,046;%%

\bibitem{Whalley:2005nh}
M. R. Whalley, D. Bourilkov and R.C. Group,
\newblock  hep-ph/0508110.
%%CITATION = HEP-PH/0508110;%%

\bibitem{Sjostrand:2006za}
T. Sjostrand, S. Mrenna and P. Skands,
\newblock JHEP 0605 (2006) 026, hep-ph/0603175.
%%CITATION = JHEPA,0605,026;%%

\bibitem{Sjostrand:2007gs}
T. Sjostrand, S. Mrenna and P. Skands,
\newblock Comput. Phys. Commun. 178 (2008) 852,
arXiv:0710.3820 [hep-ph].
%%CITATION = CPHCB,178,852;%%

\bibitem{Corcella:2000bw}
G. Corcella et al.,
\newblock JHEP 0101 (2001) 010, hep-ph/0011363.
%%CITATION = JHEPA,0101,010;%%

\bibitem{Bahr:2008pv}
M. Bahr et al.,
\newblock  Eur. Phys. J.  C58 (2008) 639,
arXiv:0803.0883 [hep-ph].
%%CITATION = EPHJA,C58,639;%%

\bibitem{Boos:2001cv}
E. Boos et al.,
\newblock hep-ph/0109068.
%%CITATION = HEP-PH/0109068;%%

\bibitem{Alwall:2006yp}
J. Alwall  et al.,
\newblock Comput.  Phys.  Commun. 176, (2007) 300,
hep-ph/0609017.
%%CITATION = CPHCB,176,300;%%

\bibitem{vanHameren:2007pt}
A. van Hameren,
\newblock arXiv:0710.2448 [hep-ph].
%%CITATION = 0710.2448;%%

\bibitem{'tHooft:1973jz}
G. 't Hooft,
\newblock Nucl. Phys. B72 (1974) 461.
%%CITATION = NUPHA,B72,461;%%

\bibitem{Maltoni:2002mq}
F. Maltoni, K. Paul, T. Stelzer and S. Willenbrock,
\newblock Phys. Rev. D67 (2003) 014026, hep-ph/0209271.
%%CITATION = PHRVA,D67,014026;%%

\bibitem{Catani:2001cc}
S. Catani, F. Krauss, R. Kuhn and B. R. Webber,
\newblock JHEP 0111 (2001) 063, hep-ph/0109231.
%%CITATION = JHEPA,0111,063;%%

\bibitem{Krauss:2002up}
F. Krauss,
\newblock JHEP 0208 (2002) 015, hep-ph/0205283.
%%CITATION = HEP-PH/0205283;%%

\bibitem{Schalicke:2005nv}
A. Schalicke and F. Krauss,
\newblock JHEP 0507 (2005) 018, hep-ph/0503281.
%%CITATION = HEP-PH/0503281;%%

\bibitem{Lavesson:2005xu}
N. Lavesson and L. Lonnblad,
\newblock JHEP 0507 (2005) 054, hep-ph/0503293.
%%CITATION = HEP-PH/0503293;%%

\bibitem{Mangano:2006rw}
M. L. Mangano, M. Moretti, F. Piccinini and M. Treccani,
\newblock JHEP 0701 (2007) 013, hep-ph/0611129.
%%CITATION = JHEPA,0701,013;%%

\bibitem{Alwall:2007fs}
J. Alwall et al.,
\newblock Eur. Phys. J.  C53 (2008) 473, arXiv:0706.2569 [hep-ph].
%%CITATION = EPHJA,C53,473;%%

\bibitem{Catani:1992zp}
S. Catani, Y. L. Dokshitzer and B. R. Webber,
\newblock Phys. Lett. B285 (1992) 291.
%%CITATION = PHLTA,B285,291;%%

\bibitem{Catani:1993hr}
S. Catani, Y. L. Dokshitzer, M. H. Seymour and B. R. Webber,
\newblock Nucl. Phys.  B406 (1993) 187.
%%CITATION = NUPHA,B406,187;%%

\bibitem{Argyres:1995ym}
E. N. Argyres et al.,
\newblock Phys. Lett. B358 (1995) 339, hep-ph/9507216.
%%CITATION = PHLTA,B358,339;%%

\bibitem{Beenakker:1996kn}
W. Beenakker et al.,
\newblock Nucl. Phys. B500 (1997) 255, hep-ph/9612260.
%%CITATION = NUPHA,B500,255;%%

\bibitem{Denner:2006ic}
A. Denner and S. Dittmaier,
\newblock Nucl. Phys. Proc. Suppl. 160 (2006) 22, hep-ph/0605312.
%%CITATION = HEP-PH/0605312;%%

\bibitem{Kleiss:1985gy}
R. Kleiss, W. J. Stirling and S. D. Ellis,
\newblock Comput. Phys. Commun. 40 (1986) 359.
%%CITATION = CPHCB,40,359;%%

\bibitem{Cabibbo:1963yz}
N. Cabibbo,
\newblock Phys. Rev. Lett. 10 (1963) 531.
%%CITATION = PRLTA,10,531;%%

\bibitem{Kobayashi:1973fv}
M. Kobayashi and T. Maskawa,
\newblock Prog. Theor. Phys. 49 (1973) 652.
%%CITATION = PTPKA,49,652;%%

\bibitem{Berends:2000fj}
F. A. Berends, C. G. Papadopoulos and R. Pittau,
\newblock Comput. Phys. Commun. 136 (2001) 148, hep-ph/0011031.
%%CITATION = HEP-PH/0011031;%%

\bibitem{Berends:1994xn}
F. A. Berends, R. Pittau and R. Kleiss,
\newblock Comput. Phys. Commun. 85 (1995) 437, hep-ph/9409326.
%%CITATION = HEP-PH/9409326;%%

\bibitem{Gleisberg:2003bi}
T. Gleisberg, F. Krauss, C. G. Papadopoulos, A. Schaelicke and S. Schumann,
\newblock Eur. Phys. J. C34 (2004) 173, hep-ph/0311273.
%%CITATION = EPHJA,C34,173;%%

\end{thebibliography}

\end{document}